\documentclass[notitlepage,nofootinbib,10pt,pra,twocolumn]{revtex4-1}
\usepackage[colorlinks=true]{hyperref}
\usepackage{amsmath}
\usepackage{amssymb}
\usepackage{color}
\definecolor{deepblue}{rgb}{0,0,0.5}
\definecolor{deepred}{rgb}{0.6,0,0}
\definecolor{deepgreen}{rgb}{0,0.5,0}
\usepackage[mmddyy]{datetime}
\usepackage{wrapfig}
\usepackage{moreverb}
\usepackage{graphicx}
\usepackage{hhline}
\usepackage{listings}
\usepackage{ctable}
\usepackage{moreverb}

\newcommand\pythonstyle{\lstset{
language=Python,
basicstyle=\ttm,
numbers=left,
stepnumber=1,
otherkeywords={self},             
keywordstyle=\ttb\color{deepblue},
emph={MyClass,__init__},          
emphstyle=\ttb\color{deepred},    
stringstyle=\color{deepgreen},
frame=tb,                         
showstringspaces=false            %
}}

\lstnewenvironment{python}[1][]
{
\pythonstyle
\lstset{basicstyle=\small,#1}
}
{}

\newcommand\pythoninline[1]{{\pythonstyle\lstinline!#1!}}

\begin{document}

\newcommand{\comb}{\mathrm{comb}}
\newcommand{\rect}{\mathrm{rect}}
\newcommand{\tri}{\mathrm{tri}}
\newcommand{\sinc}{\mathrm{sinc}}
\newcommand{\Gaus}{\mathrm{Gaus}}
\newcommand{\somb}{\mathrm{somb}}
\newcommand{\fstop}{f/\#}
\newcommand{\warn}[1]{{\color{red} \textbf{#1}}}
\newcommand{\blue}[1]{{\color{blue} #1}}
\newcommand{\Ito}{It$\hat{\mathrm{o}}$}

\title{Tree tensor network classifiers for machine learning: from quantum-inspired to quantum-assisted}
\author{Michael L.~Wall}
\email{Michael.Wall@jhuapl.edu}
\author{Giuseppe D'Aguanno}
\email{Giuseppe.DAguanno@jhuapl.edu}
\address{The Johns Hopkins University Applied Physics Laboratory, 11100 Johns Hopkins Rd. Laurel, MD 20723, USA}

\begin{abstract}

We describe a quantum-assisted machine learning (QAML) method in which multivariate data is encoded into quantum states in a Hilbert space whose dimension is exponentially large in the length of the data vector.  Learning in this space occurs through applying a low-depth quantum circuit with a tree tensor network (TTN) topology, which acts as an unsupervised feature extractor to identify the most relevant quantum states in a data-driven fashion, analogous to coarse-graining strategies used in renormalization group methodologies from statistical physics.  This unsupervised feature extractor then feeds a supervised linear classifier together with a set of truth labels for the data type, and encodes the output in a small-dimensional quantum register.  In contrast to previous work on \emph{quantum-inspired} TTN classifiers, in which the embedding map and class decision weights did not map the data to well-defined quantum states, we present an approach that can be implemented on gate-based quantum computing devices.  In particular, we identify an embedding map with accuracy similar to the recently defined exponential machines (Novikov \emph{et al.}, arXiv:1605.03795), but which produces valid quantum state embeddings of classical data vectors, and utilize manifold-based gradient optimization schemes to produce isometric operations mapping quantum states to a register of qubits defining a class decision.  We detail methods for efficiently obtaining one-point and two-point correlation functions of the vectors defining the decision boundary of the quantum model, which can be used for model interpretability, as well as methods for obtaining classification decisions from partial data vectors.  Further, we show that the use of isometric tensors can significantly aid in the human interpretability of the correlation functions extracted from the decision weights, and may produce models that are less susceptible to adversarial perturbations.  We demonstrate our methodologies in applications utilizing the MNIST handwritten digit dataset and a multivariate timeseries dataset of human activity recognition.
\end{abstract}

\maketitle

\section{Introduction}

Quantum computers--devices that utilize inherently quantum phenomena such as entanglement to process information--have long held interest due to their ability to enable disruptive, exponential speedups over classical devices in certain tasks, such as integer factorization~\cite{shor1999polynomial,gidney2019factor} and simulation of quantum chemical and materials science problems~\cite{aspuru2005simulated,bauer2020quantum}.  While scaling quantum computing devices to the millions of qubits required for reliable execution of these tasks~\cite{campbell2017roads} remains a long-term goal, we have entered an era of noisy, intermediate-scale quantum (NISQ) devices~\cite{preskill2018quantum} with tens to hundreds of qubits, noise levels too large to enable error correction, and limitations in hardware connectivity and allowed gate sets.  The NISQ computing landscape has grown beyond  the academic laboratory into an industrial reality~\cite{smith2016practical,steiger2018projectq,haner2018software,Qiskit,larose2019overview}, with several general-purpose research-scale machines available through cloud services and other specialized machines engineered for a specific demonstration~\cite{arute2019quantum}.  While the ultimate goal of NISQ devices will be to mature quantum hardware technology and algorithms towards the goal of universal, error-corrected quantum computing, a parallel effort to discover impactful near-term applications of NISQ devices is the focus of significant current research~\cite{arute2020hartree}.

Quantum-assisted machine learning (QAML), in which a quantum system forms part of a model used for statistical inferences whose fidelity is improved by interactions with data, has emerged as a promising possible avenue for NISQ devices~\cite{biamonte2017quantum,perdomo2018opportunities,ciliberto2018quantum}.  There are many key reasons for this: measurements on quantum systems have an underlying statistical interpretation, and well-performing ML models should be robust against noise, which may include the hardware noise present in NISQ devices.  Several recent results have given theoretical support for enhancements of QAML models versus classical models on certain tasks~\cite{glasser2019expressive,sweke2020quantum,coyle2020born,tangpanitanon2020expressibility}.  In the research described herein, we focus on QAML models built upon tensor networks, an enabling technology that has transformed quantum condensed matter and many-body physics in the last thirty years~\cite{schollwock2011density,orus2014practical,orus2019tensor}.  Tensor networks offer a robust framework for defining QAML models with several additional advantages: they can be executed on classical or quantum devices with an exponentially improving expressibility on quantum devices~\cite{glasser2019expressive}, and certain tensor network topologies enable sequential preparation schemes~\cite{schon2005sequential,schon2007sequential,perez2007matrix} that are highly quantum resource efficient~\cite{huggins2019towards}, a key consideration for near-term devices.

\begin{figure*}[t]
  \begin{center}
\includegraphics[width=1.8\columnwidth]{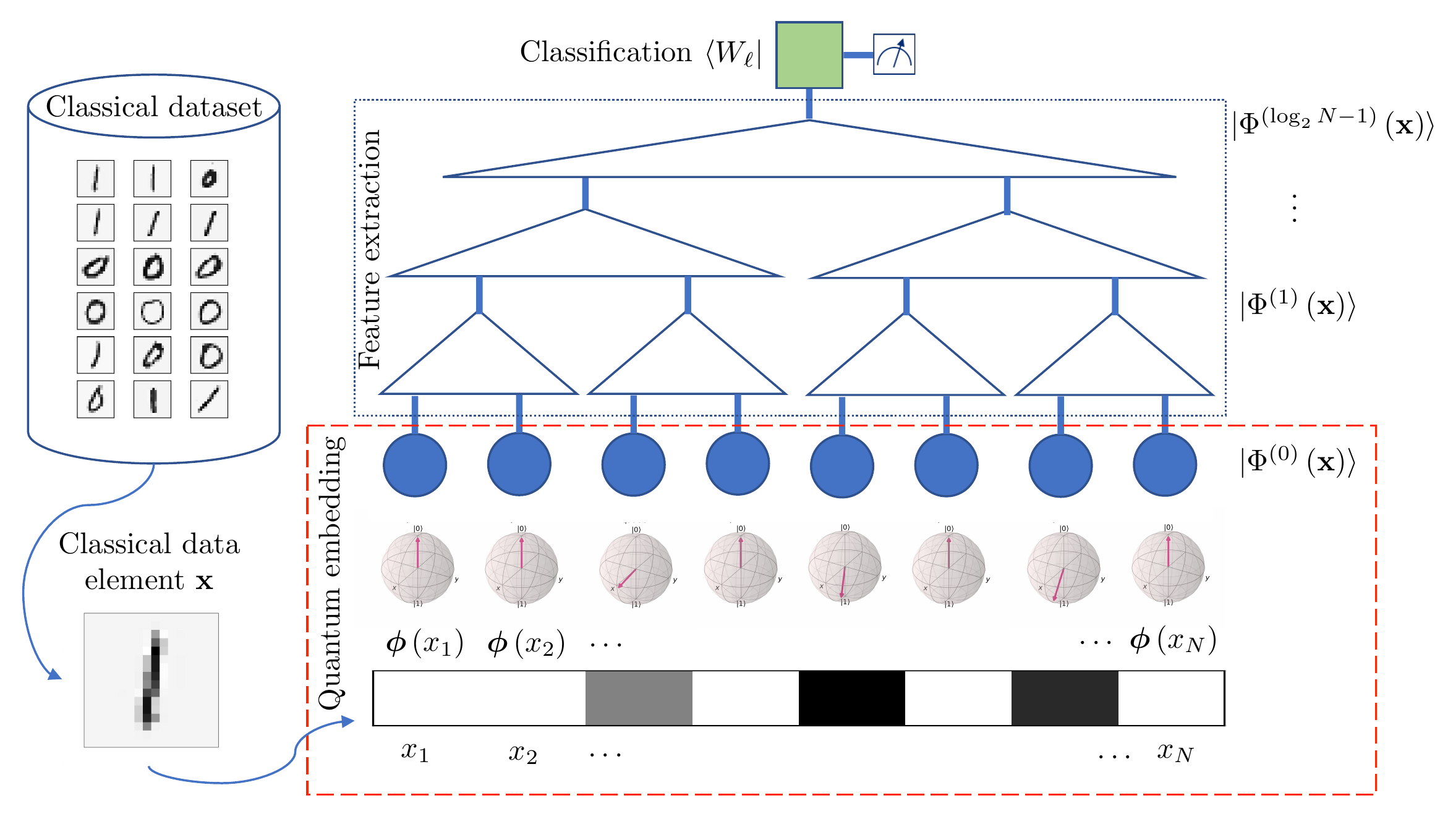}  
\caption{\label{fig:RGFig} (Color online) \emph{Overview of the machine learning workflow for a tree tensor network (TTN) classifier.}  A data instance $\mathbf{x}$ from a classical dataset is transformed into a quantum state $|\Phi^{(0)}\left(\mathbf{x}\right)\rangle$ by feeding each element of the data vector $x_i$ through a local map $\boldsymbol{\phi}\left(x_i\right)$ defining a qubit superposition, as shown schematically on the Bloch sphere.  The isometric tensors of the TTN (triangles) define a coarse graining of collections of these quantum feature vectors at progressively higher levels of scale.  At the highest level of scale, the projection of the extracted feature vector onto a collection of weight vectors $\langle W_{\ell}|$ defines a classification decision.}
\end{center}
\end{figure*}

Given the above favorable characteristics of tensor networks (TNs) as machine learning models, it is no surprise that a significant and growing body of work exists on TN-based machine learning~\cite{stoudenmire2016supervised,stoudenmire2018learning,grant2018hierarchical,PhysRevE.98.042114,carrasquilla2019reconstructing,evenbly2019number,klus2019tensor,PhysRevB.99.155131, liu2019machine,glasser2020probabilistic,trenti2020quantum, bradley2020modeling, gillman2020tensor,miller2020tensor,selvan2020tensor,wang2020anomaly,reyes2020multi,PhysRevX.8.031012, efthymiou2019tensornetwork,biamonte2018quantum,PhysRevA.102.012415,wall2020Generative,blagoveschensky2020deep,mugel2020dynamic}.  While many of these works are quantum inspired~\cite{cichocki2014tensor,cichocki2017tensor,oseledets2011tensor} in the sense that the objects appearing in the TN description are not required to correspond to physically realizable quantum states, some proposals deal with truly quantum data structures, and some have tested TN-based approaches on NISQ hardware~\cite{grant2018hierarchical,bhatia2019matrix,wall2020Generative}.  In the present work, we compare models with a tree tensor network (TTN) structure for classification that are constrained to be true quantum data structures with those that are unconstrained, using metrics of performance and interpretability.  

In defining a fully quantum approach to TTN-based classification, we introduce several innovations.  First, we identify an embedding map for translating classical data vectors into vectors in a high-dimensional Hilbert space, inspired by the classical approach of Ref.~\cite{novikov2016exponential} which efficiently maps the data to a high-order polynomial, that provides comparable performance while also producing valid quantum states.  We also discuss means for optimizing supervised learning weights mapping a collection of quantum feature vectors in a high-dimensional space to a small register of class decision qubits.  Namely, we utilize gradient descent algorithms on the Riemannian manifold of isometric tensors such that this mapping between Hilbert spaces corresponds to an allowed transformation between orthonormal quantum states.  We show that these techniques, together with the quantum interpretation of the resulting model structure, enable novel applications such as classification of partial data vectors and interpretability analyses.  Results are given for the canonical Modified National Institute of Standards and Technology (MNIST) handwritten digit dataset~\cite{lecun2010mnist}, in which we investigate both the classification of all digits zero through nine as well as the simpler problem of distinguishing zeros from ones, as well as a time series dataset using smartphone accelerometer and gyroscope data to recognize human activity.


\section{Tree tensor networks for quantum-assisted machine learning}
\label{sec:TTNML}

Tensor networks (TNs) represent the high-rank tensor expressing the quantum wavefunction of a multi-partite system as a contraction over low-rank tensors, and hence define families of low-rank approximations whose computational power can be expressed in terms of the maximum dimension of any contracted index $\chi$, known as the bond dimension.  A wide variety of TN topologies have been considered which are able to efficiently capture certain classes of quantum states~\cite{schollwock2011density,orus2014practical,orus2019tensor}; in the present work we focus on tree tensor networks (TTNs)~\cite{shi2006classical}, which utilize the tree topology exemplified in Fig.~\ref{fig:RGFig}.  Here and throughout, we will utilize the Penrose graphical notation for tensors~\cite{schollwock2011density}, in which a symbol represents a tensor, the number of lines coming off from a tensor is its rank, and when a line connects two tensors summation over that common index is implied.  A symbol with lines pointing upwards represents a tensor, and a symbol with the lines pointing downwards represents its complex conjugate.

A key feature of TTN models for interpretation is that they are representations of real-space \emph{renormalization groups} (RGs).  RGs mathematically formalize the changes in the description of a physical system as we change the scale at which it is observed, e.g. length scale or energy scale~\cite{wilson1983renormalization}, and include a procedure for keeping the most relevant degrees of freedom when performing this coarse graining transformation.  In statistical physics it is observed that the behavior of many physical systems can be grouped into a small number of universality classes that depend on only a few parameters when viewed from very coarse scales; these universality classes may correspond to phases of matter, for example.  Similarly, in machine learning we want to extract features from data at all scales in order to classify into broad categories specified by only a few relevant variables; tensor networks are a means of formalizing this approach.  Based on this analogy, we will refer to the vertical layers of tensors in a TTN as levels of scale, indexed by the superscript $\xi$ of $|\Phi^{(\xi)}\left(\mathbf{x}\right)\rangle$ as in Fig.~\ref{fig:RGFig}.

We now turn to formalizing the procedure for using a TTN for quantum-assisted machine learning of classical data.  Broadly, our process for QAML has the key elements of (1) encoding of classical data into a quantum state, (2) learning of a quantum model (tensor network) on encoded training data, and (3) readout of quantum model results.  Using these elements, we can also choose between QAML models with different hyperparameters (e.g., the bond dimension of the TN) using separate validation data, and also test the ultimate performance of these models on a held-out test set.  In the subsections to follow, we flesh out each of the items in the workflow in more detail.

\subsection{Mapping of classical data to quantum states}
\label{sec:encoding}
In this section, we address the encoding of a classical data vector $\mathbf{x}\in\mathbb{R}^L$ into a quantum state, depicted visually by the dashed, red-bordered box in Fig.~\ref{fig:RGFig}~\cite{lloyd2020quantum,schuld2020effect,PhysRevA.102.032420}.  We will denote the encoded state at the lowest level of scale as $|\Phi^{\left(0\right)}\left(\mathbf{x}\right)\rangle$.  We will consider that each element $x_j$ of the data vector is encoded into a $d$-dimensional quantum state ($d=2$ for a qubit), such that the full wavefunction $|\Phi^{\left(0\right)}\left(\mathbf{x}\right)\rangle$ lives in a $d^L$-dimensional Hilbert space.  The only restriction we will place on the encoding of classical data in quantum states is that each classical data vector is encoded in an unentangled product state, denoted by the absence of horizontal connecting lines in Fig.~\ref{fig:RGFig}.  Utilizing unentangled states has several advantages: they are the simplest to prepare experimentally with high fidelity and also enable us to use qubit-efficient sequential preparation schemes~\cite{schon2005sequential,schon2007sequential,perez2007matrix,huggins2019towards}.  Encoding individual data vectors in product states also ensures that any entanglement in the quantum ML model arises from correlations in an ensemble of data and not from \emph{a priori} assumptions about pre-existing correlations for individual data vectors~\cite{stoudenmire2016supervised}.  We will parameterize the map from an $L$-dimensional classical data vector $\mathbf{x}$ to an ensemble of $L$ $2$-level systems (qubits) as
\begin{align}
\label{eq:fullmap} |\Phi^{\left(0\right)}\left(\mathbf{x}\right)\rangle&=\bigotimes_{j=1}^{L} \left(\sum_{i_j=1}^{2}\phi^{(j)}_{i_j}\left(x_j\right)|i_j\rangle\right)\, .
\end{align}
That is, the parameterization is accomplished in terms of local maps $\boldsymbol{\phi}^{(j)}\left(x\right)$ mapping a single data element into a superposition of qubit states (see Fig.~\ref{fig:RGFig}).  In order to ensure that the full map $\Phi^{\left(0\right)}\left(\mathbf{x}\right)$ maps each data instance into a normalized vector in Hilbert space, we require that 
\begin{align}
\label{eq:Nmap} \sum_i \left|\phi_i^{(j)}\left({x}\right)\right|^2=1\;\;\; \forall x\, .
\end{align}
This condition is satisfied by the phase-like encoding
\begin{align}
\label{eq:phasemap}\phi_0\left(x\right)&=\cos\left(\frac{\pi}{2}\frac{x-x_{\mathrm{min}}}{x_{\mathrm{max}}-x_{\mathrm{min}}}\right)\, ,\\
\nonumber \phi_1\left(x\right)&=\sin\left(\frac{\pi}{2}\frac{x-x_{\mathrm{min}}}{x_{\mathrm{max}}-x_{\mathrm{min}}}\right)\, ,
\end{align}
that has been used in Refs.~\cite{stoudenmire2016supervised,liu2019machine,roberts2019tensornetwork,efthymiou2019tensornetwork} to encode data for quantum-inspired ML applications, and is shown schematically by the Bloch sphere representations of each classical data element in Fig.~\ref{fig:RGFig}.  In Ref.~\cite{novikov2016exponential} a quantum-inspired algorithm using MPSs (known in the numerical analysis community as tensor trains~\cite{oseledets2011tensor}) found good performance in certain learning tasks using the map
\begin{align}
\label{eq:PolynomialMap}\phi_0\left(x\right)&=1\, ,\;\; \phi_1\left(x\right)=ax\, .
\end{align}
This has the appealing property that each data vector $\mathbf{x}$ is mapped into a weighted superposition of all correlations $x_ix_j\dots x_k$ of orders 0 through $L$ with each $x_i$ appearing at most once, and so a tensor network weighting vector in this space can efficiently select out the most relevant correlations from this exponentially large set.  The factor $a$ can be used to scale the data and avoid numerical overflow.  Unfortunately, this map does not satisfy the condition in Eq.~\eqref{eq:Nmap}, and so is not suitable for application on quantum hardware.  We can define a generalization of Eq.~\eqref{eq:phasemap} as
\begin{align}
\label{eq:ScaledPhaseMap}\phi_0\left(x\right)&=\cos\left(a\frac{x}{x_{\mathrm{max}}}\right)\, ,\;\; \phi_1\left(x\right)=\sin\left(a\frac{x}{x_{\mathrm{max}}}\right)\, ,
\end{align}
with $a\sim\mathcal{O}\left(0.1\right)$, which satisfies Eq.~\eqref{eq:Nmap} and so is suitable for quantum discriminative applications.  Noting that $a\frac{x}{x_{\mathrm{max}}}\ll 1$ $\forall x$, we can use the small-angle approximation to find
\begin{align}
\phi_0\left(x\right)&=1+\mathcal{O}\left(a^2\right)\, ,\;\; \phi_1\left(x\right)=a\frac{x}{x_{\mathrm{max}}}+\mathcal{O}\left(a^3\right)\, ,
\end{align}
and so this map has the same essential features as Eq.~\eqref{eq:PolynomialMap}.  A comparison of model performance utilizing the maps Eq.~\eqref{eq:ScaledPhaseMap} and Eq.~\eqref{eq:PolynomialMap} will be given in Sec.~\ref{sec:App}.

\subsection{Construction of unsupervised feature extractor}

We now turn to building an unsupervised feature extractor as a TTN acting on the encoded quantum data description, following Ref.~\cite{stoudenmire2018learning}.  This procedure is depicted graphically by the dotted blue box in Fig.~\ref{fig:RGFig}.  We assume that we have a collection of $M$ training data vectors $\{\mathbf{x}_m\}$, $m=1,\dots,M$, that has been encoded into a collection of training states $\{|\Phi\left(\mathbf{x}_m\right)\rangle\}$.  Using this map and these states, we would like to build a model $f\left(\mathbf{x}\right)=\langle W|\Phi\left(\mathbf{x}\right)\rangle$ parameterized by a weight vector $|W\rangle$ in the many-body Hilbert space, to perform some machine learning task (e.g., classification).  Given our collection of training vectors, the optimal weights for a given task can be represented by a linear combination of the training vectors in the quantum space as $\langle W|=\sum_m \alpha_m\langle \Phi\left(\mathbf{x}_m\right)|$ where the $\{\alpha_m\}$ are task-specific.  This is the \emph{representer theorem}, which states that the weights lie within the span of feature vectors generated by the training data.  If the effective dimension of the set of relevant feature vectors is much smaller than the full dimension of the space, this will enable us to efficiently compress our model representation.  Ignoring issues of efficiency, one way to find the effective dimension of this set would be from the singular value decomposition (SVD) of the matrix $\Phi_{\mathbf{i} m}\equiv \langle i_1\dots i_L|\Phi^{\left(0\right)}\left(\mathbf{x}_m\right)\rangle$, where $\mathbf{i}=\left(i_1,\dots,i_L\right)$ is a multi-index spanning the full $d^L$ dimensional Hilbert space.  Namely, given $\Phi_{\mathbf{i} m}\to \sum_{\mu} U_{\mathbf{i}\mu}S_{\mu} V_{\mu m}$, we have $W_{\mathbf{i}}=\sum_{\mu} \beta_{\mu} U_{\mathbf{i}\mu}$.  The matrix $\Phi_{\mathbf{i} m}$ is $d^L\times M$ dimensional, and so for typical training dataset sizes $M\gtrsim 10^4$ a direct decomposition of this matrix is unfeasible.

Instead of the direct approach using the SVD, we will instead utilize an iterative approach that builds a TTN to extract the relevant feature vectors.  We start by noting that the SVD matrix $U$, whose columns form a basis for the statistically significant set of feature vectors from the training dataset, can alternately be obtained as the eigenvectors of the feature space covariance matrix
\begin{align}
\rho_{\mathbf{ii}'}&=\frac{1}{M}\sum_{m=1}^{M} \Phi_{\mathbf{i} m}\Phi_{\mathbf{i}' m}=\sum_{\mu} U_{\mathbf{i}\mu}S_{\mu}^2U_{\mathbf{i}'\mu}^{\star}\, .
\end{align}
Translating back to the language of quantum mechanics, we see that this object is nothing but the density operator obtained from an incoherent sum of all training vectors 
\begin{align}
\label{eq:fullrho} \hat{\rho}&=\frac{1}{M}\sum_m|\Phi\left(\mathbf{x}_m\right)\rangle\langle \Phi\left(\mathbf{x}_m\right)|\, .
\end{align}
Recall that each of our feature vectors $|\Phi^{\left(0\right)}\left(\mathbf{x}_m\right)\rangle$ at the lowest level of scale is unentangled according to the restrictions placed in Sec.~\ref{sec:encoding}; however, the sum over feature vectors introduces correlations that may manifest themselves as entanglement in feature space.  If we look at only a subset of the data elements in the feature space, correlations with other elements will induce fluctuations in this subset that are captured as impurity in the corresponding reduced density operator.  The tree tensor network construction will capture these correlations in feature space hierarchically, using a procedure motivated by how the dominant fluctuations manifest themselves in certain quantum mechanical systems~\cite{stoudenmire2018learning}.

\begin{figure*}[t]
  \begin{center}
\includegraphics[width=1.99\columnwidth]{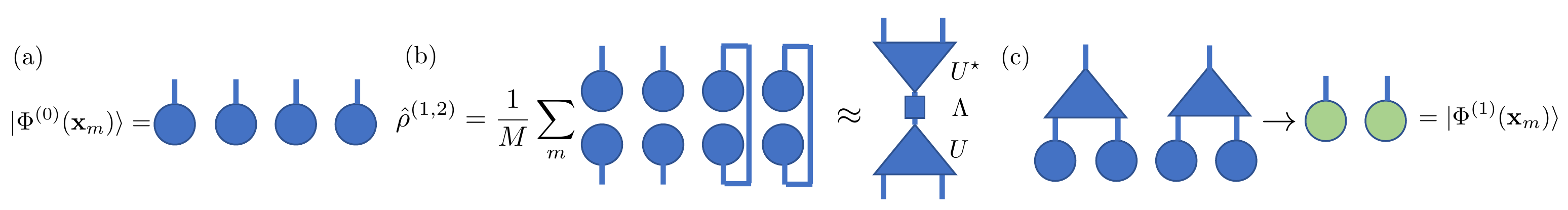}  
\caption{\label{fig:TNR} (Color online) \emph{Tensor network diagrams} for (a) mapping of a data vector to a product state in Hilbert space (b) approximation of a two-site reduced density operator in terms of an isometric tensor, and (c) renormalization of a data vector using the TTN isometries, resulting in a new product state in the renormalized Hilbert space.}
\end{center}
\end{figure*}

We begin the TTN construction procedure by defining a set of operations which group pairs of data elements together.  We will construct these operations to capture as much of the subsystem fluctuations as possible given restricted resources, which will be defined explicitly below.  We do so by constructing the reduced density operators of pairs of neighboring sites\footnote{Note that while we index the data elements using $j=1,\dots,L$, this is simply a numbering scheme and does not require a one-dimensional topology for the data space.}
\begin{widetext}
\begin{align}
\nonumber \hat{\rho}^{(2j-1,2j)}=&\frac{1}{M}\sum_{m=1}^{M}\sum_{i_{2j-1}i_{2j-1}'i_{2j}i_{2j}'} \Phi^{(0)}_{i_{2j-1}}\left(x_{m,2j-1}\right)\Phi^{(0)}_{i_{2j-1}'}\left(x_{m,2j-1}\right)\Phi^{(0)}_{i_{2j}}\left(x_{m,2j}\right)\Phi^{(0)}_{i_{2j}'}\left(x_{m,2j}\right)\\
&\times |i_{2j-1}i_{2j}\rangle\langle i_{2j-1}'i_{2j}'|\prod_{j'\ne 2j-1,2j}\left(\sum_{i=1}^{2}\left|\Phi^{(0)}_{i}\left(x_{m,j'}\right)\right|^2\right)\, .
\end{align}
\end{widetext}
This represents the effective covariance of the $(2j-1)$ and $(2j)$ elements of the data vector in feature space, taking into account the average effect of correlations with all other elements of the data vector.  Just as with the full density operator Eq.~\eqref{eq:fullrho}, these reduced density operators are Hermitian and so admit an eigendecomposition
\begin{align}
\hat{\rho}^{(2j-1,2j)}_{(i_{2j-1}i_{2j})(i_{2j-1}'i_{2j}')}&=\sum_{\mu_j}U_{(i_{2j-1}i_{2j})\mu_j}^{(0)[j]\star}\Lambda_{\mu_j}U_{(i_{2j-1}'i_{2j}')\mu_j}^{(0)[j]}\, ,
\end{align}
denoted graphically in Fig.~\ref{fig:TNR}(b).  The eigenvectors, specified as columns of the matrices $\mathbb{U}^{(0)[j]}$, define a change of basis from the ``bare" feature space $|i_{2j-1}i_{2j}\rangle$ to a ``renormalized" feature space $|\mu_j\rangle$ as 
\begin{align}
|\mu_j\rangle&=\sum_{i_{2j-1}i_{2j}}U^{(0)[j]}_{(i_{2j-1}i_{2j})\mu_j}|i_{2j-1}i_{2j}\rangle\, .
\end{align}
In general, $\hat{\rho}^{(2j-1,2j)}$ has $d^2$ nonzero eigenvalues, and so a complete change of basis will require all of the eigenvectors to fully reproduce the training data.  However, in many physically relevant cases, not all possible feature vectors are seen, or some combinations occur with negligibly low probability.  In this case, we can keep only a restricted number $\chi$ of the eigenvectors corresponding to the largest eigenvalues, and hence define an \emph{isometric} transformation from the original $d^2$-dimensional space to a $\chi$-dimensional space.  Alternatively, we can obtain $\chi$ implicitly by requiring that we discard at most $\varepsilon$ of the Frobenius norm of the reduced density operator through
\begin{align}
\label{eq:varepsilondef} 1-(\sum_{\alpha=1}^{\chi} \rho_{\alpha\alpha})/(\sum_{\alpha}\rho_{\alpha\alpha})\le \varepsilon\, .
\end{align}
The strategy of using the states corresponding to the largest eigenvalues of a reduced density operator to renormalize the space is known as \emph{White's rule}, and is the underpinning of the celebrated density matrix renormalization group (DMRG) method in quantum physics~\cite{white1992density,schollwock2011density}.

Collecting up all of the isometric tensors defined from applying the procedure above to a complete, disjoint set of data element pairs, we have defined an isometric transformation to a globally renormalized feature space
\begin{align}
\langle \boldsymbol{\mu}|\Phi^{(1)}\left(\mathbf{x}\right)\rangle&=\sum_{\mathbf{i}}\left(\prod_{j=1}^{L/2}U_{(i_{2j-1},i_{2j})\mu_j}^{(0)[j]}\right)\langle \mathbf{i}|\Phi^{(0)}\left(\mathbf{x}\right)\rangle \, ,
\end{align}
as shown graphically in Fig.~\ref{fig:TNR}(c).  Importantly, the renormalized feature space still has a product structure between its component vectors, even though the components are now of higher dimension ($\chi$ instead of $d$), and account for correlations between subcomponents at a lower scale.  Hence, we can iterate this renormalization procedure up until the highest scale.  At the highest level of scale, we are left with a single index $\mu$ labeling the relevant set of feature vectors for the dataset.  The entire hierarchical network, when all intermediate bond indices have been contracted, defines an isometry approximately corresponding to the eigendecomposition of the full density operator Eq.~\eqref{eq:fullrho}.  The network does not have to be contracted up to the highest level of scale to be useful; stopping before the highest level of scale maps the original high-dimensional space into another multi-partite quantum state representation.  Refs.~\cite{stoudenmire2018learning,reyes2020multi} have investigated models in which a TN with a tree-like structure are augmented by another TN structure, such as a matrix product state.  Finally, we note that this procedure defines a family of models indexed by the maximum bond dimensions $\chi$ of the isometries (alternatively, through the norm cutoff $\varepsilon$ defined in Eq.~\eqref{eq:varepsilondef}) which form the main hyperparameter of the models.  This cutoff can be used as a regularization to avoid overfitting.

\subsection{Training of a supervised classifier}
\label{sec:training}

So far in the QAML workflow we have encoded our classical data instances into quantum states in a $d^L$-dimensional quantum Hilbert space, and then utilized a tree tensor network feature extractor that acts as an isometry projecting these quantum feature vectors onto a small, relevant set.  The isometric feature extractor was constructed in an unsupervised manner, utilizing the features extracted from the training data and White's rule for determining which features were the most relevant.  Now that we have extracted the relevant features, we turn to optimizing a mapping from the extracted features to a classification decision, denoted by the green box at the top of the network in Fig.~\ref{fig:RGFig}.  Let us again consider that we have training data $\{\mathbf{x}_m\}$, $m=1,\dots,M$ that are in one of $\ell=1,\dots,C$ classes, with the truth class label of data vector $m$ denoted as $\ell_m$.  If we denoted the extracted feature vector at the highest level of scale as $|\Phi\left(\mathbf{x}\right)\rangle$, we now wish to define weights $\langle W_{\ell}|$ such that the functions $f_{\ell}\left(\mathbf{x}\right)=\langle W_{\ell}|\Phi\left(\mathbf{x}\right)\rangle$ are 1 when $\mathbf{x}$ is in class $\ell$ and zero otherwise.  For simplicity, we do so by optimizing an unregularized quadratic cost function~\cite{stoudenmire2018learning}
\begin{align}
\nonumber \mathcal{C}&=\frac{1}{2M}\sum_{m=1}^{M}\sum_{\ell=1}^{C}\left(f_{\ell}\left(\mathbf{x}_m\right)-\delta_{\ell\ell_m}\right)^2\, ,\\
\label{eq:quadcost}&=\frac{1}{2M}\sum_{m=1}^{M}\sum_{\ell=1}^{C}\left(\langle W_{\ell}|\Phi\left(\mathbf{x}_m\right)\rangle-\delta_{\ell\ell_m}\right)^2\, .
\end{align}
The least-squares solution that optimizes each weight vector independently is
\begin{align}
\label{eq:W}W_{\nu,\ell}&=\left[\langle \nu|\Phi\left(\mathbf{x}_m\right)\rangle\right]^{-1}\boldsymbol{\delta}_{\ell\ell_m}\, ,
\end{align}
in which $\left[\bullet\right]^{-1}$ denotes the Moore-Penrose pseudoinverse and $\boldsymbol{\delta}_{\ell\ell_m}$ is a length $M$ vector that is 1 when $\ell_m=\ell$ and zero otherwise.  With these optimized weight vectors in hand, we can classify a test vector $\mathbf{x}_{\mathrm{test}}$ according to $\mathrm{argmax}_{\ell}|\langle W_{\ell}|\Phi\left(\mathbf{x}_{\mathrm{test}}\right)\rangle |^2$.  As above, if the renormalization procedure is stopped before the highest level of scale, then the weight vector $\langle W_{\ell}|$ is a multipartite tensor mapping the degrees of freedom remaining after renormalization into a class decision label $\ell$.

While the above pseudoinverse solution is optimal from the perspective of the cost function, the resulting matrix $\mathbb{W}$ is not isometric and so the set of vectors $W_{\ell}$ do not constitute orthonormal quantum states in the coarse-grained feature space.  We remedy this here by considering an optimization procedure that seeks to minimize the cost function Eq.~\eqref{eq:quadcost} directly within the space of isometric matrices.  Complex isometric matrices of dimension $\chi\times C$ (with $\chi\ge C$) define a Riemannian manifold known as the \emph{Stiefel manifold}~\cite{edelman1998geometry}
\begin{align}
\mathrm{St}\left(\chi,C\right)=\left\{\mathbb{W}\in \mathbb{C}^{\chi\times C} : \mathbb{W}^{\dagger}\mathbb{W}=\mathbb{I}\right\}\, .
\end{align}
Several recent works~\cite{hauru2020riemannian,luchnikov2020riemannian} have discussed the application of manifold optimization techniques to isometric tensors for quantum applications, which we briefly review here to keep the exposition self-contained.  First, we will define a few tools from Riemannian geometry that will be useful, the first being the orthogonal projection of a general complex matrix $\mathbb{D}\in\mathbb{C}^{\chi\times C}$ onto the tangent space $\mathcal{T}_{\mathbb{W}}\mathcal{M}$ of the manifold $\mathcal{M}=\mathrm{St}\left(\chi,C\right)$ at the point $\mathbb{W}$.  For the Stiefel manifold, this projection takes the form
\begin{align}
\mathcal{P}_{\mathbb{W}}\left(\mathbb{D}\right)&=\mathbb{D}-\frac{1}{2}\mathbb{W}\left(\mathbb{W}^{\dagger}\mathbb{D}+\mathbb{D}^{\dagger}\mathbb{W}\right)\, .
\end{align}
The next operation we will need is retraction, which is a smooth map $\mathcal{R}_{\mathbb{W}}\left(\bullet\right)$ from $\mathcal{T}_{\mathbb{W}}\mathcal{M}\to\mathcal{M}$ that offers a computationally efficient alternative to the exponential map generalizing point transformations in Euclidean space.  Retractions are not unique, but can be defined by any smooth map satisfying the conditions
\begin{align}
\mathcal{R}_{\mathbb{W}}\left(\mathbf{0}\right)&=\mathbb{W}\, ,\;\;\forall \mathbb{W}\in \mathcal{M}\, ,\\
\frac{d}{dt}\mathcal{R}_{\mathbb{W}}\left(t\mathbf{v}\right)|_{t=0}&=\mathbf{v}\,, \;\;\forall \mathbf{v}\in\mathcal{T}_{\mathbb{W}}\mathcal{M}\, ,
\end{align} 
in which $\mathbf{0}$ denotes the zero vector in $\mathcal{T}_{\mathbb{W}}\mathcal{M}$.  In practice, we utilize the SVD retraction defined by
\begin{align}
\mathcal{R}_{\mathbb{W}}\left(\boldsymbol{\eta}\right)&=\mathbb{UV}\, ,\;\; \left(\mathbb{W}+\boldsymbol{\eta}\right)=\mathbb{U}\Sigma\mathbb{V}\, ,
\end{align}
where the rightmost equality defines the SVD.  The final definition we need is vector transport, which is a computationally efficient alternative to parallel transport which generalizes transport of a point $x$ along a vector direction $v$ from Euclidean geometry.  As with retraction, vector transport is not unique, but is any smooth map $\tau_{\mathbb{W}}\left(\mathbf{y},\boldsymbol{\eta}\right)$, $\mathbb{W}\in\mathcal{M}$, $\mathbf{y},\boldsymbol{\eta}\in\mathcal{T}_{\mathbb{W}}\mathcal{M}$ with the following properties
\begin{align}
\tau_{\mathbb{W}}\left(\mathbf{y},\boldsymbol{\eta}\right)&\in \mathcal{T}_{\mathcal{R}_{\mathbb{W}}\left(\boldsymbol{\eta}\right)}\mathcal{M}\, , \\ 
\tau_{\mathbb{W}}\left(\mathbf{y},\mathbf{0}\right)&=\mathbf{y}\, ,\\
\tau_{\mathbb{W}}\left(a\mathbf{y}+b\mathbf{z},\boldsymbol{\eta}\right)&=a\tau_{\mathbb{W}}\left(\mathbf{y},\boldsymbol{\eta}\right)+b\tau_{\mathbb{W}}\left(\mathbf{z},\boldsymbol{\eta}\right)\, .
\end{align}
In the results given below, we use the vector transport
\begin{align}
\tau_{\mathbb{W}}\left(\mathbf{y},\boldsymbol{\eta}\right)&=\mathcal{P}_{\mathcal{R}_{\mathbb{W}}\left(\boldsymbol{\eta}\right)}\left(\mathbf{y}\right)\, ,
\end{align}
with the SVD retraction defined above.

With the above definitions in hand, we can now define a gradient optimization procedure directly within the space of isometric matrices.  We recall the standard iterative scheme for gradient descent of a cost function $\mathcal{C}\left(\mathbf{w}\right)$ with momentum parameter $\beta$ and learning rate $\eta$:
\begin{align}
\mathbf{m}_{i+1}&=\beta\mathbf{m}+\left(1-\beta\right)\nabla \mathcal{C}\left(\mathbf{w}_i\right)\, ,\\
\mathbf{w}_{i+1}&=\mathbf{w}_i-\eta\mathbf{m}_{i+1}\, .
\end{align}
Here, $i$ is the iteration index.  We can generalize this to manifold gradient descent as
\begin{align}
\label{eq:MGD1}\tilde{\mathbf{m}}_{i+1}&=\beta\mathbf{m}_i+\left(1-\beta\right)\mathcal{P}_{\mathbb{W}_i}\left[\nabla \mathcal{C}\left(\mathbb{W}_i\right)\right]\, ,\\
\mathbb{W}_{i+1}&=\mathcal{R}_{\mathbb{W}_i}\left(-\eta \tilde{\mathbf{m}}_{i+1}\right)\, ,\\
\label{eq:MGD3}\mathbf{m}_{i+1}&=\tau_{\mathbb{W}_{i}}\left(\tilde{\mathbf{m}}_{i+1},-\eta\tilde{\mathbf{m}}_{i+1}-\eta\beta\left(\tilde{\mathbf{m}}_{i+1}-\mathbf{m}_{i}\right)\right)\, .
\end{align}
For the quadratic cost function defined in Eq.~\eqref{eq:quadcost}, we have that
\begin{align}
&\left[\nabla \mathcal{C}\right]_{\nu\ell}=2\frac{\partial \mathcal{C}}{\partial W^{\star}_{\nu\ell}}\, ,\\
&=\frac{1}{M}\sum_{m=1}^{M}\langle \nu |\Phi\left(\mathbf{x}_m\right)\rangle\left(\langle \Phi\left(\mathbf{x}_m\right)|W_{\ell}\rangle-\delta_{\ell\ell_m}\right)\, .
\end{align}
A convenient starting point to initialize the isometric optimization procedure is the nearest isometric matrix to the pseudoinverse solution $\mathbb{W}_{\mathrm{pinv}}$ in the $L_2$-norm, given by
\begin{align}
\label{eq:Wisonearest} \mathbb{W}_{\mathrm{iso}}&=\mathbb{UV}\, , \mathbb{W}_{\mathrm{pinv}}=\mathbb{U}\Sigma \mathbb{V}\, .
\end{align}
Example cost function behavior during optimization will be given later in Sec.~\ref{sec:App}.

\subsection{Application of classifier on test data}
\label{sec:test}

\begin{figure*}[t]
  \begin{center}
\includegraphics[width=1.3\columnwidth]{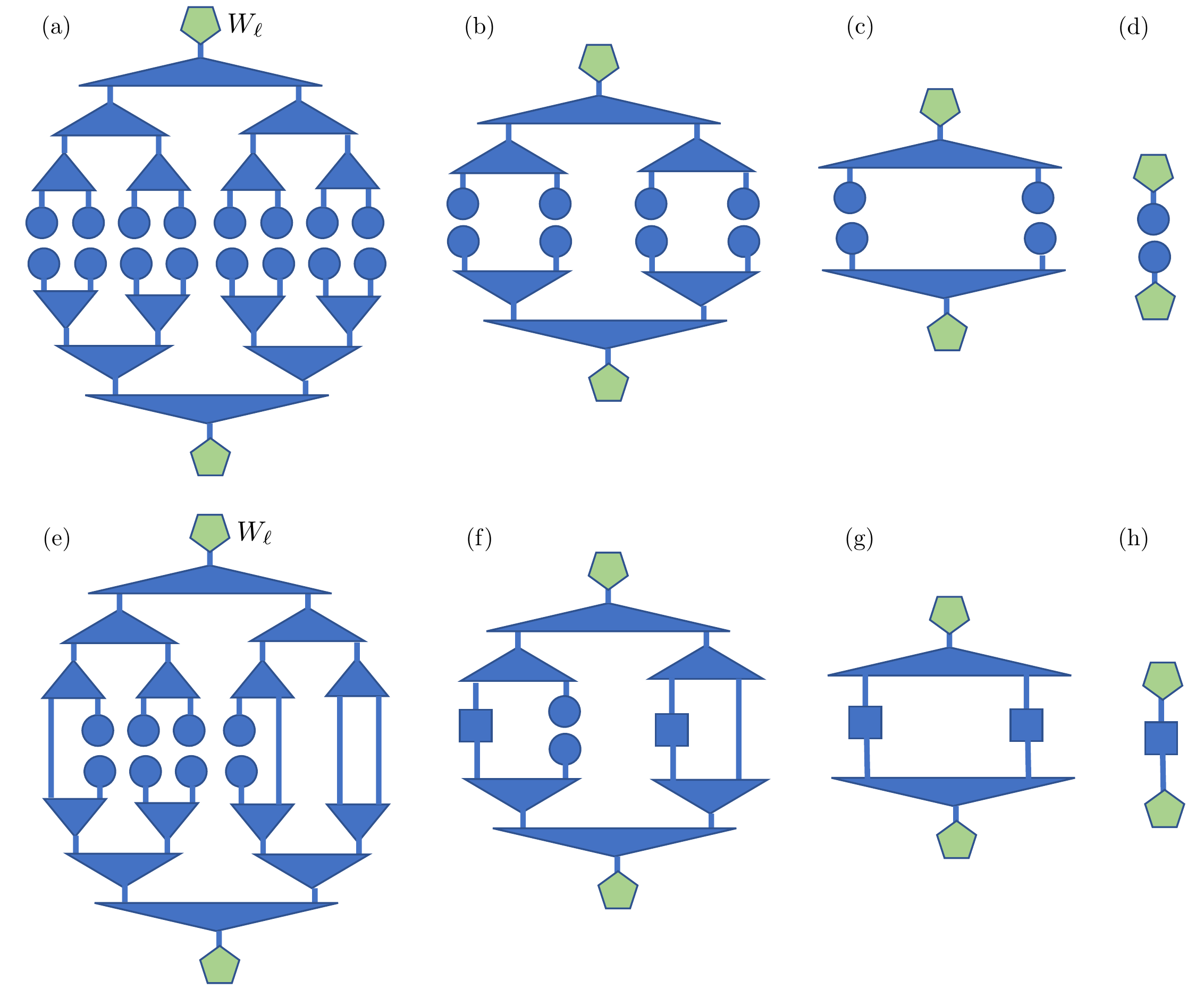}  
\caption{\label{fig:FullAndPartial} (Color online) \emph{Classification decision function evaluation with complete and partial data.} Panels (a)-(d): evaluation of a TTN model with a full data vector amounts to renormalization of feature vectors followed by an inner product.  Panels (e)-(h): evaluation of a TTN model with partial information requires renormalization not only of feature vectors, but also of density operators (squares).}
\end{center}
\end{figure*}

At this stage we have a complete pipeline to go from a classical data instance to a class decision: the data instance $\mathbf{x}$ is embedded into a product state feature vector using the map $\boldsymbol{\phi}\left(x_j\right)$ resulting in the state $|\Phi^{(0)}\left(\mathbf{x}\right)\rangle$, this state is coarse grained using the TTN feature extractor into the state $|\Phi^{(\xi)}\rangle$ at the level of scale $\xi$, and then the inner product of this coarse grained feature vector with the weights $\mathbb{W}$ defines a decision function $\mathrm{argmax}_{\ell}|\langle W_{\ell}|\Phi\left(\mathbf{x}_{\mathrm{test}}\right)\rangle |^2$ mapping to a class label $\ell$.  If the test data has the same dimension as the training data (i.e.~there are no missing elements in the data vector), the tensor network used to evaluate the decision function amounts to the recursive renormalization of feature vectors followed by an inner product, as shown schematically in Fig.~\ref{fig:FullAndPartial}(a)-(d).  In this section, we broaden the applicability of this procedure to include cases in which data is missing for certain elements of the test vector.  Given that we are dealing with quantum data structures, a lack of subsystem information in our approach is naturally handled through the formalism of density operators.  However, the renormalization procedure for density operators differs from the of feature vectors, and will be explained herein.

For the case in which our test vector contains partial information, the tensor network to be contracted for the decision functions takes a form as in Fig.~\ref{fig:FullAndPartial}(e)-(h), in which the dimensions with missing data are contracted.  We recall that the tensors $U^{(\xi)[j]}_{(ii')\mu_j}$ are isometric, and so obey the row-orthogonality condition $\sum_{ii'}U^{(\xi)[j]}_{(ii')\mu_j}U^{(\xi)[j]\star}_{(ii')\mu_j'}=\delta_{\mu_j\mu_j'}$.  Hence, if no data is present in any dimension of a tensor at level $\xi$, the trace is simply shifted to next level of scale, as seen in going from panel (e) to panel (f) of Fig.~\ref{fig:FullAndPartial}.  If one of the elements of $U$, say the element corresponding to index $i$, contains a feature vector, while the other element $i'$ contains no information, the result is not another feature vector, but instead a density operator, denoted by a square in Fig.~\ref{fig:FullAndPartial}.  Renormalization of any object together with a density operator produces another density operator.  This calculus of renormalization can be summarized with a collection of rules $\left(\bullet,\bullet\right)\to \mathcal{O}$, in which $\left(\bullet,\bullet\right)$ denotes the two objects input to an isometric tensor and $\mathcal{O}$ denotes its output.  Namely, the rules are
\begin{align}
\nonumber &\left(|,|\right)\to |\, ,\;\; \left(\circ,\circ\right)\to \circ\, ,\;\; \left(|,\circ\right)\to \Box\, ,\\
&\left(\circ,|\right)\to \Box\, ,\;\; \left(\Box,\ast\right)\to \Box\, \,\;\; \left(\ast,\Box\right)\to \Box\, ,
\end{align}
in which $|$ denotes a contracted index (no data), $\circ$ denotes a feature vector, $\Box$ denotes a density operator, and $\ast$ denotes any object.  From an implementation standpoint, the order in which indices are contracted is essential for obtaining the best scaling algorithm~\cite{schollwock2011density}.  In particular, it is advisable to sum over at most one contracted index at a time, storing the results in intermediate tensors, in order to reduce the total number of operations.  Further, density operators obtained as intermediaries will not generally be of full rank, and so can be eigendecomposed and the resulting set of feature vectors renormalized and reconstructed more efficiently than direct renormalization of the density operator.

As seen in Fig.~\ref{fig:FullAndPartial}(e)-(h), when utilizing partial information to extract a class decision, the relevant quantity can be written as $\mathrm{argmax}_{\ell}\langle W_{\ell}|\hat{\rho}\left(\mathbf{x}_{\mathrm{test}}\right)|W_{\ell}\rangle$, in which $\hat{\rho}\left(\mathbf{x}_{\mathrm{test}}\right)$ results from the renormalization of $\mathrm{Tr}_{\mathcal{S}}|\Phi^{(0)}\left(\mathbf{x}_{\mathrm{test}}\right)\rangle\langle \Phi^{(0)}\left(\mathbf{x}_{\mathrm{test}}\right)|$.  Here, $\mathcal{S}$ is the set of all indices where $\mathbf{x}_{\mathrm{test}}$ does not have support.  We find that when a significant portion of the data is absent, the resulting density matrix $\hat{\rho}$ is highly mixed.  Hence, the class decision function will be a sum over many terms $\mathrm{argmax}_{\ell}\left[\sum_{\lambda} p_{\lambda}\left| \langle W_{\ell}|\lambda\rangle\right|^2\right]$, in which $p_{\lambda}$ are the eigenvalues of $\hat{\rho}$ and $|\lambda\rangle$ the associated eigenvectors.  We have found that the best performance is observed when we instead project the weight vectors onto only the eigenvector with highest probability, i.e.~$\mathrm{argmax}_{\ell}\left[ p_{1}\left| \langle W_{\ell}|1\rangle\right|^2\right]$, at least for the present training strategy which utilizes only the full data vectors.  More details on performance will be presented in Sec.~\ref{sec:App}.

\subsection{Interpretability analysis}
\label{sec:interp}

\begin{figure*}[t]
  \begin{center}
\includegraphics[width=1.9\columnwidth]{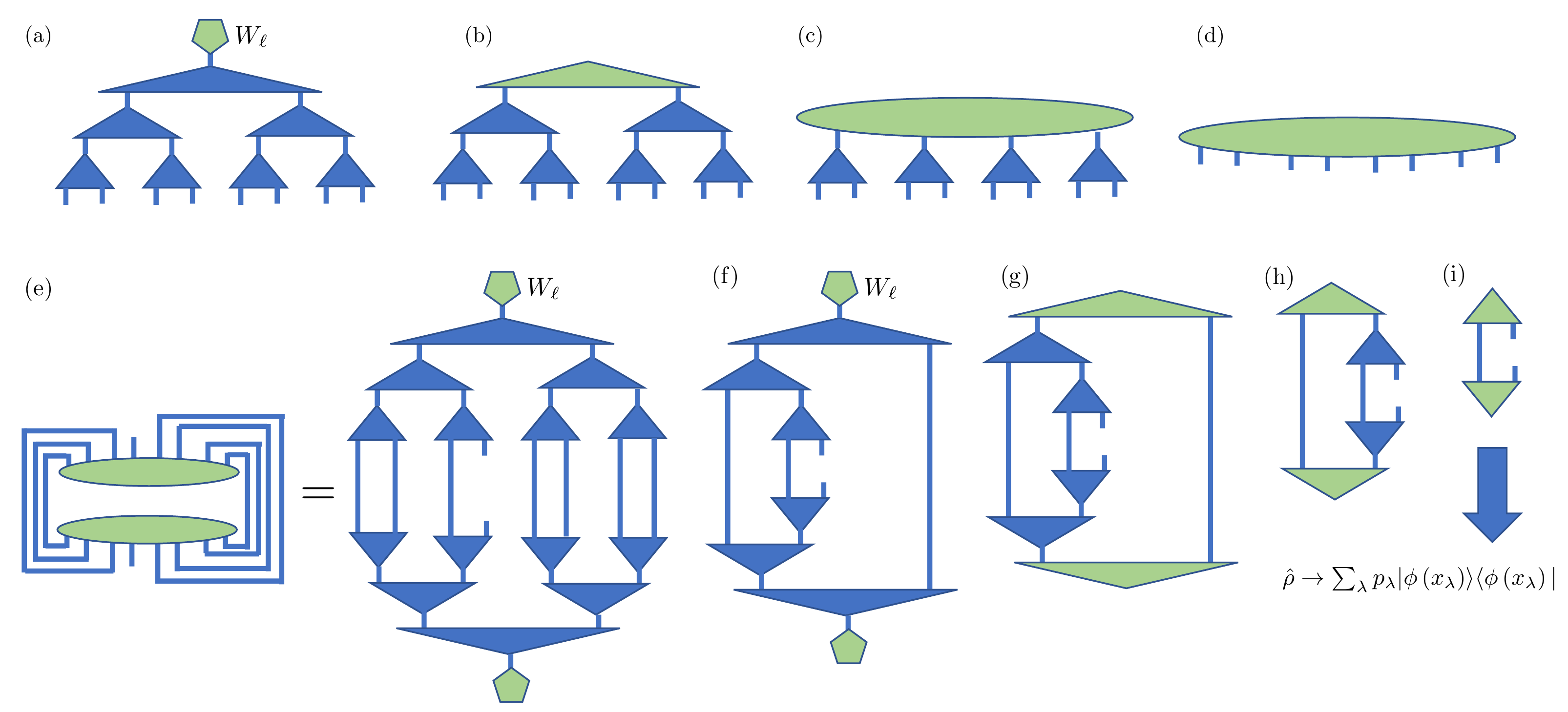}  
\caption{\label{fig:OneBodyRho} (Color online) \emph{Fine-graining of feature vectors from higher to lower levels of scale.} Panels (a)-(d): ``Fine graining" of a classification decision vector to a many-body state at the raw data scale.  Panels (e)-(i): Looking at a single data-scale element of the fine-grained decision vector can be performed efficiently using the TTN structure.}
\end{center}
\end{figure*}

In this section, we highlight a property of the TTN feature extractor which can aid in interpretability of the classification decision.  Namely, since the TTN is formed through isometric tensors, it is also possible to read the tensor network in reverse and so ``fine grain" a decision vector from the highest level of scale back to the lowest, ``data" level of scale.  This is likely a more general feature of quantum machine learning algorithms, as the unitary nature of the operations involved means that the computation should be reversible, taking as input a classification decision vector and producing a vector at the feature scale.  In principle, this fine graining proceeds as in Fig.~\ref{fig:OneBodyRho} (a)-(d).  However, the resulting vector lives in the full $d^L$-dimensional Hilbert space, and so cannot be immediately interpreted.  A natural way of analyzing this state is to look at low-order correlation functions or moments, analogous to looking at the magnetization or magnetic susceptibility of a many-body quantum spin system~\cite{altland2010condensed}.  In panels (e)-(i) of Fig.~\ref{fig:OneBodyRho}, we see that tracing over all data vector elements but one produces a ``one-point" reduced density matrix $\hat{\rho}^{(1)}_{ii'}$ that can be efficiently obtained using the tensor network structure of the feature extractor.  From the one-point reduced density matrix $\hat{\rho}^{(1)}_{ii'}$, we can interpret its eigenvectors in terms of the embedding map $\boldsymbol{\phi}\left(x\right)$ as in Eq.~\eqref{eq:fullmap}, and so ``undo" the original feature space embedding back into the classical data space.  As an example, for the scaled phase map Eq.~\eqref{eq:ScaledPhaseMap}, we find the average data element from the eigendecomposition $\hat{\rho}^{(1)}\to\{p_{\lambda},|\lambda\rangle\}$ as
\begin{align}
\label{eq:xinterp} \langle x\rangle/x_{\mathrm{max}}&=\sum_{\lambda}p_{\lambda} \mathrm{tan}^{-1}\left(\frac{\langle 1|\lambda\rangle}{\langle 0|\lambda\rangle}\right)/a\, .
\end{align}

Interpretations of the classification weights vectors can also be extended to higher-order correlations~\cite{trenti2020quantum}.  As an example, we can construct the two-point density matrix $\hat{\rho}^{(2)}_{\left(ii'\right),\left(jj'\right)}$ using diagrams analogous to Fig.~\ref{fig:OneBodyRho} (e)-(i).  In order to isolate the contributions from correlations that do not appear in the lowest-order moment, it is useful to a define the correlation density matrix~\cite{cheong2009correlation,munder2010correlation} as
\begin{align}
\label{eq:CDM}\hat{\rho}^{(C)}_{\left(ii'\right),\left(jj'\right)}&=\hat{\rho}^{(2)}_{\left(ii'\right),\left(jj'\right)}-\hat{\rho}^{(1)}_{ii'}\hat{\rho}^{(1)}_{jj'}\, .
\end{align}
From a quantum perspective, the correlation density matrix is useful because its trace with any product operator provides the associated connected correlation function
\begin{align}
\mathrm{Tr}\left[\hat{\rho}^{(C)} \hat{O}_i\otimes \hat{O}_j\right]&=\langle \hat{O}_i\hat{O}_j\rangle-\langle \hat{O}_i\rangle\langle \hat{O}_j\rangle\, .
\end{align}
Matrix decompositions of the correlation density matrix can be used to identify the dominant correlations present in a quantum system for detailed analysis, as discussed in Refs.~\cite{cheong2009correlation,munder2010correlation}.  A related scalar metric of the degree of correlation between sites $i$ and $j$ is given by the mutual information~\cite{PhysRevLett.100.070502}
\begin{align}
\label{eq:MI}I\left(i,j\right)&=H\left(\hat{\rho}^{(1)}_i\right)+H\left(\hat{\rho}^{(1)}_j\right)-H\left(\hat{\rho}^{(2)}_{ij}\right)\, ,
\end{align}
in which $H\left(\hat{\rho}\right)=-\mathrm{Tr}\left[\hat{\rho}\ln\hat{\rho}\right]$ is the von Neumann entropy.  Examples will be provided along with the applications in Sec.~\ref{sec:App}.

\section{Compilation for a quantum computer}

So far, our discussion has been focused on the training and deployment of a quantum-inspired machine learning model on a classical computer.  However, one of the key features of tensor networks is that they correspond to low-depth quantum circuits, and so provide a natural ``blueprint" for compilation onto a quantum device~\cite{huggins2019towards,schon2005sequential}.  In this section, we will provide some detail into a formal process for taking a classically trained TTN model and converting it into a sequence of quantum operations to be applied on a quantum computer.  We will refer to this process as \emph{quantum compilation}.  Our process is formal because it does not account for the limited connectivity and native gate sets of present-day quantum computers, nor does it optimize the number or type of operations to account for realistic machine noise.  We expect that similar techniques to those applied to matrix product state generative models~\cite{wall2020Generative}, in which the inherent freedom of the TN structure is utilized to aid in compiling to quantum hardware and greedy heuristics are used to compile isometries into native operations on target hardware, can be utilized for TTN models for classification.  This stage of quantum compilation enables optimal classical optimization strategies to be utilized to ``precondition" a quantum model, defining a model architecture and initial guesses at parameters for the gate set and topology of a given device.  This architecture can then be further refined directly on the quantum device using a hybrid quantum/classical optimization strategy~\cite{PhysRevA.99.032331,schuld2020circuit,bergholm2018pennylane,sweke2020stochastic,broughton2020tensorflow}.

The key operation in the TTN feature extractor is the application of an isometry $U_{\mu_{\xi}\nu_{\xi}}^{\mu_{\xi+1}}$ at the level of scale $\xi$ to two feature vectors $|\phi_{\mu_{\xi}}\rangle$ and $|\psi_{\nu_{\xi}}\rangle$ to produce a new feature vector $|\Xi_{\xi_{\mu}+1}\rangle$ at the level of scale $(\xi+1)$.  Assuming that the states at level $\xi$ can be described by a $\chi_{\xi}$-dimensional basis of states, they can each be encoded into a quantum register of $\log_2\chi_{\xi}$ qubits.  Hence, each is an operation on $2\log_2\chi_{\xi}$ qubits that produces a result on $\log_2\chi_{\xi+1}$ qubits with the remainder being decoupled so that they can be re-initialized and hence re-utilized in the computation.  Using such operations leads to a resource requirement of $\mathcal{O}\left(\log_2L\log_2\chi\right)$ qubits to apply a bond dimension $\chi$ TTN feature extractor to an $L$-dimensional data vector encoded to a quantum state as in Sec.~\ref{sec:encoding}, provided the quantum device allows for mid-circuit measurement and reuse (MCMR).  This scaling demonstrates that TTN models can be extremely quantum resource efficient, and the scaling with $\chi$ further indicates that TTN models on quantum hardware can be significantly more expressive than on classical hardware~\cite{huggins2019towards}.

The operation that is specified through the TTN optimization is isometric, but all operations applied in a gate-based quantum computer are required to be unitary.  One simple means of defining a unitary operation from an isometry is to specify the desired state of the $\left(2\log_2\chi_{\xi}-\log_2\chi_{\xi+1}\right)$ decoupled qubits, call it $|\Upsilon_{\zeta}\rangle=\sum_{\zeta}\Upsilon_{\zeta}|\zeta\rangle$, and then to use the unitary generalization of the orthogonal Procrustes problem applied to the matrix $\left[\mathcal{U}\right]_{\left(\mu_{\xi+1}\zeta\right)\left(\mu_{\xi}\nu_{\xi}\right)}=\Upsilon_{\zeta}U_{\mu_{\xi}\nu_{\xi}}^{\mu_{\xi+1}}|\mu_{\xi+1}\rangle|\zeta\rangle\langle\mu_{\xi}\nu_{\xi}|$, in which $\left(a,b\right)$ denotes the Kronecker product of the indices $a$ and $b$.  The unitary matrix closest to $\mathcal{U}$ in the $L_2$-norm is given by $UV$, in which $U\Sigma V$ is the singular value decomposition of $\mathcal{U}$.  This unitary operator is still formal in the sense that it must be compiled into the native set of gates and topology for given target hardware.  Alternatively, variational schemes which measure a distance functional between a family of unitaries with optimizable parameters and the isometry, similar to those developed in Ref.~\cite{wall2020Generative} for matrix product state models, can be employed, which can remove the requirement to specify the state of the decoupled qubits.  Such methods can be devised to operate in a hardware-aware fashion where the unitary ansatz is built from allowed gates for a given hardware.

In the case that the weight vectors $\{|W_{\ell}\rangle\}$ discussed in Sec.~\ref{sec:training} define an isometric matrix, the operation that takes the feature vectors output from the TTN extractor into a collection of $\log_2 C$ class decision qubits is of the same form as those described in the previous paragraph, and so the same methods can be applied.  Following application of this operation, the class decision qubits can be measured in the computational basis and the outcome is the predicted class index in a binary representation.  In the case that the weight vectors do not form an isometric matrix, we can still extract the decision functions $f_{\ell}\left(\mathbf{x}\right)=\langle W_{\ell}|\Phi\left(\mathbf{x}\right)\rangle$ using a SWAP test as follows.  We first couple an ancilla qubit to the $\chi$-dimensional state at the highest level of scale to form the state $|\psi_{\ell}\rangle=\left[|0\rangle|W_{\ell}\rangle+|1\rangle|\Phi\left(\mathbf{x}\right)\rangle\right]/\sqrt{2}$, and then measure the probability for the ancilla to be in the state $|\leftarrow\rangle =\left[|0\rangle-|1\rangle\right]/\sqrt{2}$.  This probability is $\left|\langle \leftarrow|\psi_{\ell}\rangle\right|^2=\left[1-f_{\ell}\left(\mathbf{x}\right)\right]/2$, from which the decision function can be extracted.  In the often-encountered case that $|W_{\ell}\rangle$ is not of unit norm, the normalized state $|W_{\ell}\rangle/\sqrt{\langle W_{\ell}|W_{\ell}\rangle}$ can be used to form the state $|\psi_{\ell}\rangle$, and then the norm added back in post-processing to extract $f_{\ell}\left(\mathbf{x}\right)$.  By performing $C$ such measurements, one for each value of $\ell$, all decision functions can be extracted and the results classically post-processed for the class decision.  Clearly, dealing with a set of weight vectors that does not form an orthonormal set requires more complex operations and additional overhead in quantum resources and number of circuit runs.

\section{Example applications}
\label{sec:App}

In this section, we detail applications of the above approach to two datasets.  The first is the canonical MNIST handwritten digit dataset~\cite{lecun2010mnist}, which is commonly used as a machine learning benchmark.  This dataset consists of $28\times28$ pixel grayscale arrays of the digits 0 through 9, and we consider both classification of digits into all 10 classes as well as the simpler problem of identifying the digits 0 and 1, for the purposes of illustration.  In order to show the utility of TTN-based machine learning beyond image data, we also apply our methodologies to a multivariate human activity recognition (HAR) time series dataset~\cite{anguita2013public}.  Here, data extracted from accelerometers and gyroscopes attached to test subjects are used to infer the type of activity being performed, e.g., walking or sitting still.  Our specific data pre-processing steps will be presented together with the classifier analysis in the following subsections.

\subsection{MNIST handwritten digit dataset}

\begin{figure}[t]
  \begin{center}
\includegraphics[width=0.99\columnwidth]{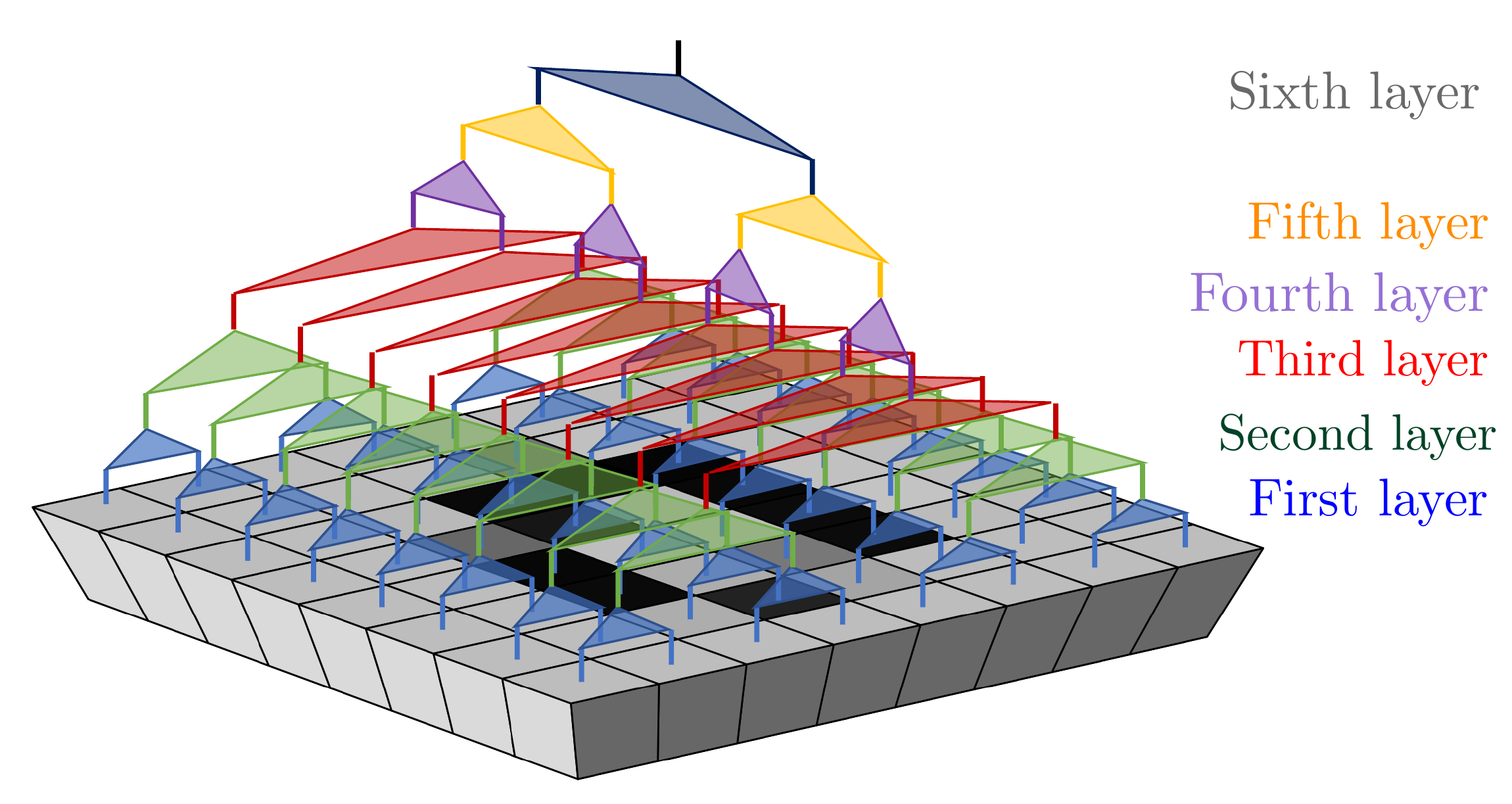}  
\caption{\label{fig:ResampTopo} (Color online) \emph{Topology of TTN renormalization for images}.  Our TTN feature extractor first renormalizes pixels neighboring within rows (first three layers), followed by renormalization of neighboring columns (last three layers).  For clarity, the procedure is shown for an image interpolated to $8\times 8$ rather than the $16\times 16$ used for classification.}
\end{center}
\end{figure}

Our first steps in utilizing the MNIST dataset are to resize and scale the data uniformly.  We utilize a bivariate spline to interpolate the data from its original $28\times28$ size into $16\times 16$ arrays for the convenience of having the total number of pixels be a power of two, and then scale the data such that each pixel in the training dataset is between zero and one.  Examples of data elements for the digits 0 and 1 processed in this fashion are shown in Fig.~\ref{fig:RGFig}.  The topology of the TTN is such that rows of the data are renormalized in lower layers and columns in higher layers, as exemplified in Fig.~\ref{fig:ResampTopo} using an $8\times8$ interpolation for clarity.  The first task that we consider using this dataset is building an unsupervised feature extractor on all 60,000 elements of the training dataset, and then defining a supervised classification into the ten classes of digit values zero through nine, tested on all 10,000 elements of the test set.  As a scalar metric of performance, we use the class-averaged $F_1$ score, which is obtained from precision $p$ and recall $r$ as 
\begin{align}
p_i &=\frac{C_{ii}}{\sum_j C_{ji}}\, ,\\
r_i &=\frac{C_{ii}}{\sum_j C_{ij}}\, ,\\
\left[F_1\right]_i&=2\frac{p_ir_i}{p_i+r_i}\, .
\end{align}
Here, $\mathbb{C}$ is the confusion matrix indexed by classes whose element $C_{ij}$ represents the number of data elements predicted to be in class $j$ whose truth class is $i$.

\begin{figure}[t]
  \begin{center}
\includegraphics[width=0.90\columnwidth]{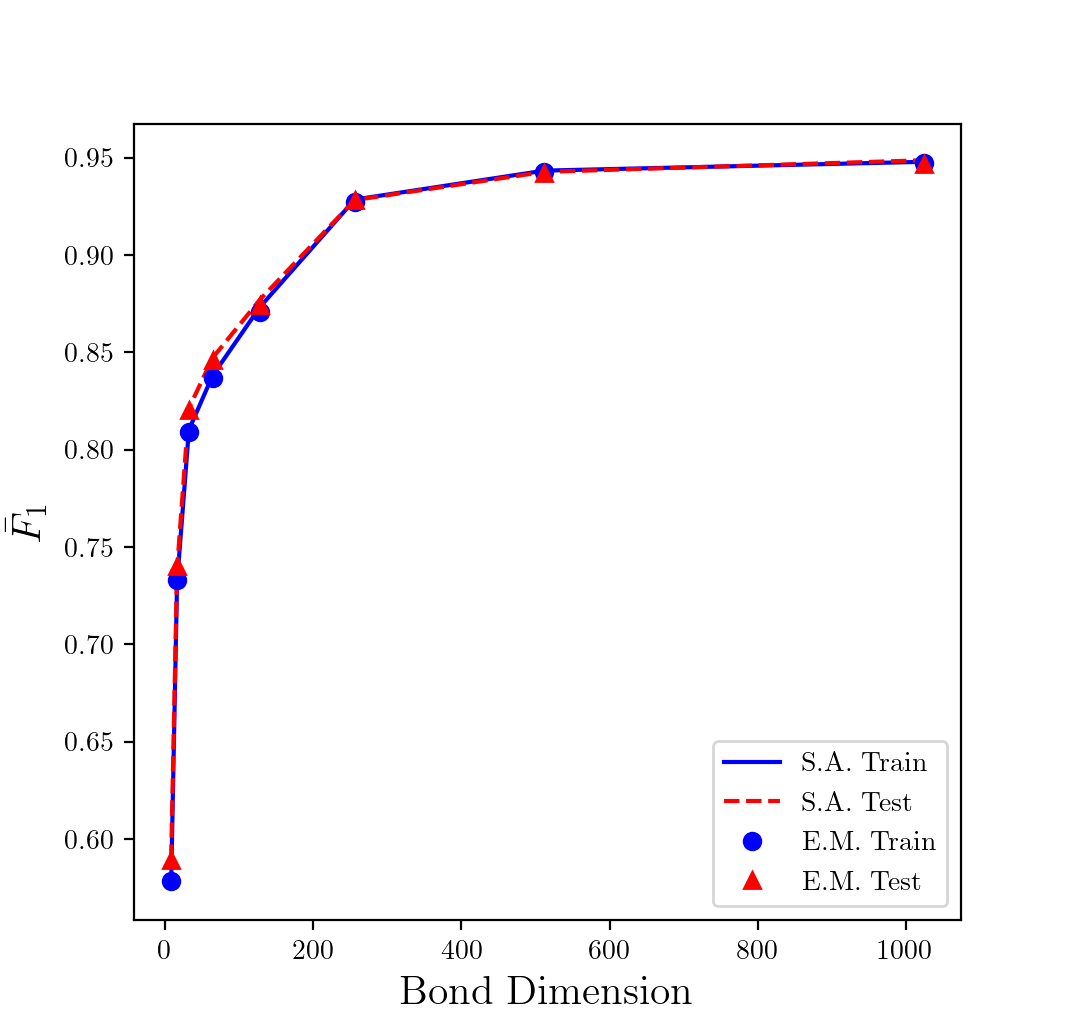}  
\caption{\label{fig:MNISTBD} (Color online) \emph{Dependence of average $F_1$ score on tensor network bond dimension}.  The average $F_1$ score on the training (test) data set for all ten digits is shown vs.~the bond dimension of the top layer with the solid blue (dashed red) line.  Lines correspond to the small-angle map Eq.~\eqref{eq:ScaledPhaseMap}, and symbols to the exponential machine map Eq.~\eqref{eq:PolynomialMap}, both with $a=0.1$.}
\end{center}
\end{figure}

We begin by comparing the average $F_1$ scores computed for the training and test sets as a function of the bond dimension of the top layer, $\chi$, in Fig.~\ref{fig:MNISTBD}.  Here, the solid blue (dashed red) lines correspond to the training (test) set average $F_1$ score using the small angle map, Eq.~\eqref{eq:ScaledPhaseMap}, with $a=0.1$.  The results for the exponential machine map~\cite{oseledets2011tensor}, Eq.~\eqref{eq:PolynomialMap}, with $a=0.1$ are shown with symbols.  We note that the unsupervised feature extraction layers were trained using a density matrix eigenvalue cutoff $\varepsilon=7\times10^{-5}$ as defined in Eq.~\eqref{eq:varepsilondef}, and then only the top layer restricted to a maximum bond dimension of $\chi$.  For small bond dimension $\chi<128$ we obtain the optimal weights by direct construction of the pseduoinverse, while for larger bond dimensions we obtain the weights using a sparse least-squares solver~\cite{fong2011lsmr}.  The results of Fig.~\ref{fig:MNISTBD} demonstrate that the small-angle and exponential machine maps have the same qualitative features from a learning perspective, and show that TTNs have sufficient expressive power to classify this dataset with high accuracy.

\begin{figure}[t]
  \begin{center}
\includegraphics[width=0.90\columnwidth]{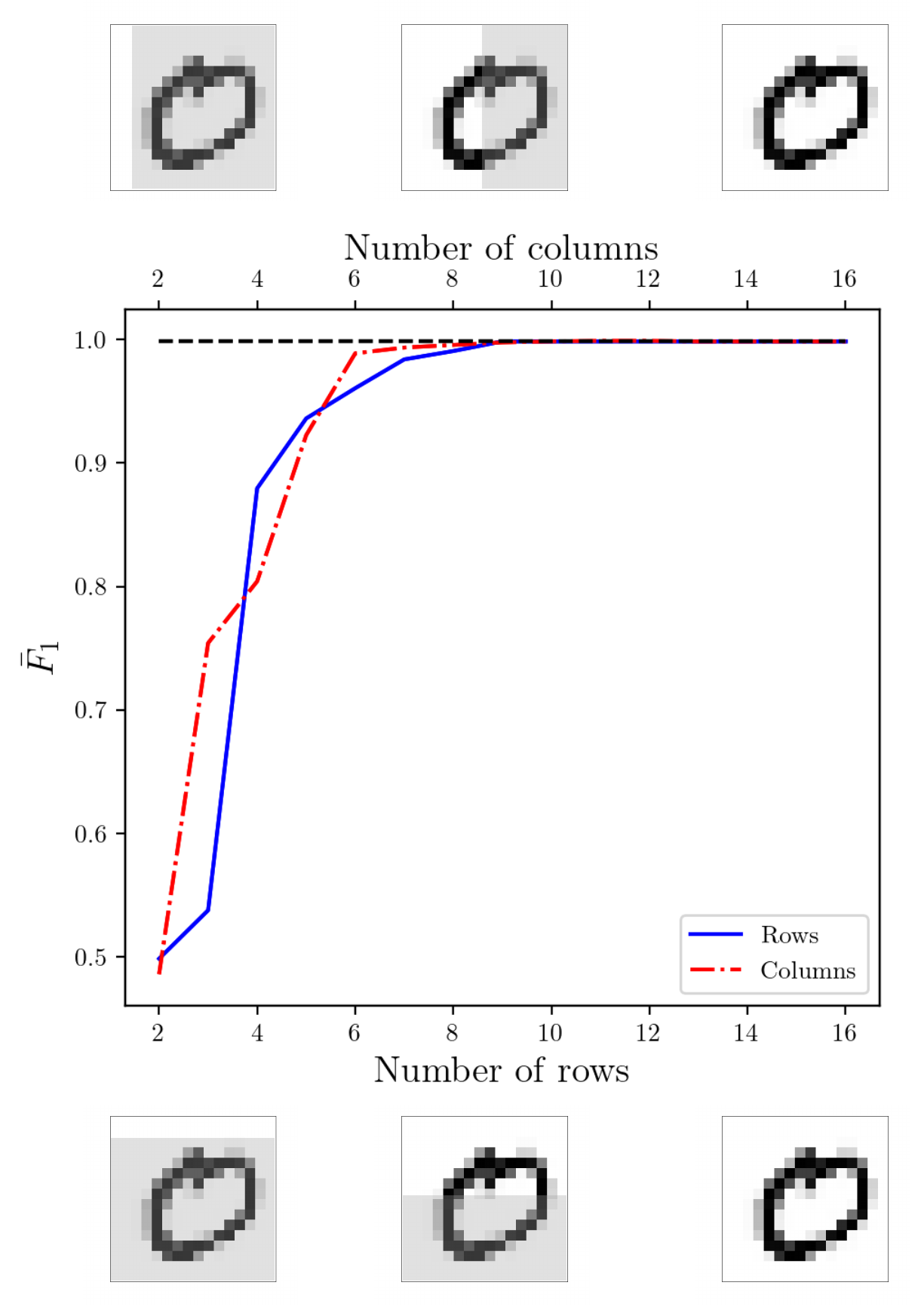}  
\caption{\label{fig:RAC} (Color online) \emph{Dependence of average test $F_1$ score on amount of data}.  The average $F_1$ score on the test data set of determining zeros vs.~ones is shown for a range of partial data.  The bottom axis corresponds to the blue solid line using a subset of the rows of an image, and the top axis corresponds to the red dot-dashed line using a subset of columns, with the ordering as indicated on an element of the test dataset near the axes.  The black dashed line corresponds to all data being used.}
\end{center}
\end{figure}

In order to enable a more detailed analysis in what follows, we now specialize to a simpler classification task of distinguishing the digits zero and one from the MNIST dataset.  We again train on resized and scaled data, now from the relevant 12,665 elements of the training set and testing on the relevant 2,115 elements of the test set, and utilize the small-angle map Eq.~\eqref{eq:ScaledPhaseMap} with $a=0.1$.  Unsurprisingly, a modest cutoff of $\varepsilon=2\times10^{-4}$ resulting in a maximum bond dimension of $79$ is sufficient to get a near-perfect average $F_1$ score of $\bar{F}_1=0.9985$ on the test dataset.  We now turn to the evaluation of the model when only partial data is utilized, as described in Sec.~\ref{sec:test}.  Fig.~\ref{fig:RAC} shows the results when only a subset of data rows or columns are utilized, with the ordering of the utilized data as shown.  The results for a single row or column are not shown due to there being a ``border" around the images that results in no useful information being present.  We see a rapid increase in $\bar{F}_1$ as more data is included until the results essentially saturate at the score given by the full data array, occurring near the point where half of the data is used for classification.  We note that more accurate results with partial data may be possible by using an alternate training strategy that utilizes a cost function involving partial data vectors.

\begin{figure}[t]
  \begin{center}
\includegraphics[width=0.85\columnwidth]{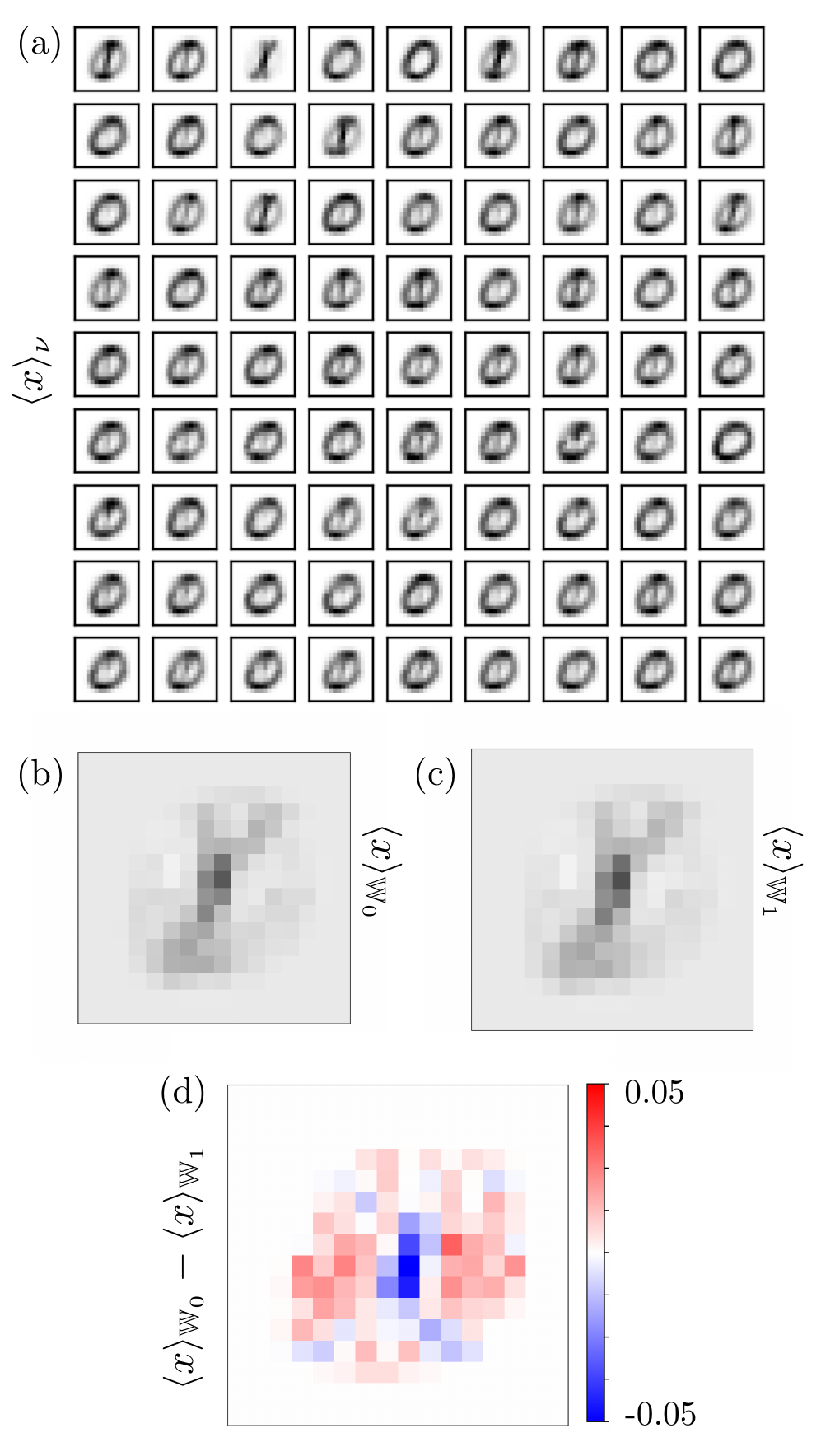}  
\caption{\label{fig:NII}  (Color online) \emph{One-point averages of classification weight vectors at the data scale}.  The data-scale representation of the one-point averages of vectors at the highest level of scale are displayed as images.  Panel (a) displays the representations of all 81 feature vectors extracted by the TTN, showing their mixed nature between 0 and 1 due to the unsupervised procedure.  Panels (b) and (c) are the data-scale representations of the weight vectors for class 0 and class 1, respectively.  While they look very similar, their difference, shown in a zoomed view in panel (d), shows that the weight vector for class 1 is higher in the central region and the weight vector for class 0 is higher towards the edges.}
\end{center}
\end{figure}

\begin{figure}[t]
  \begin{center}
\includegraphics[width=0.99\columnwidth]{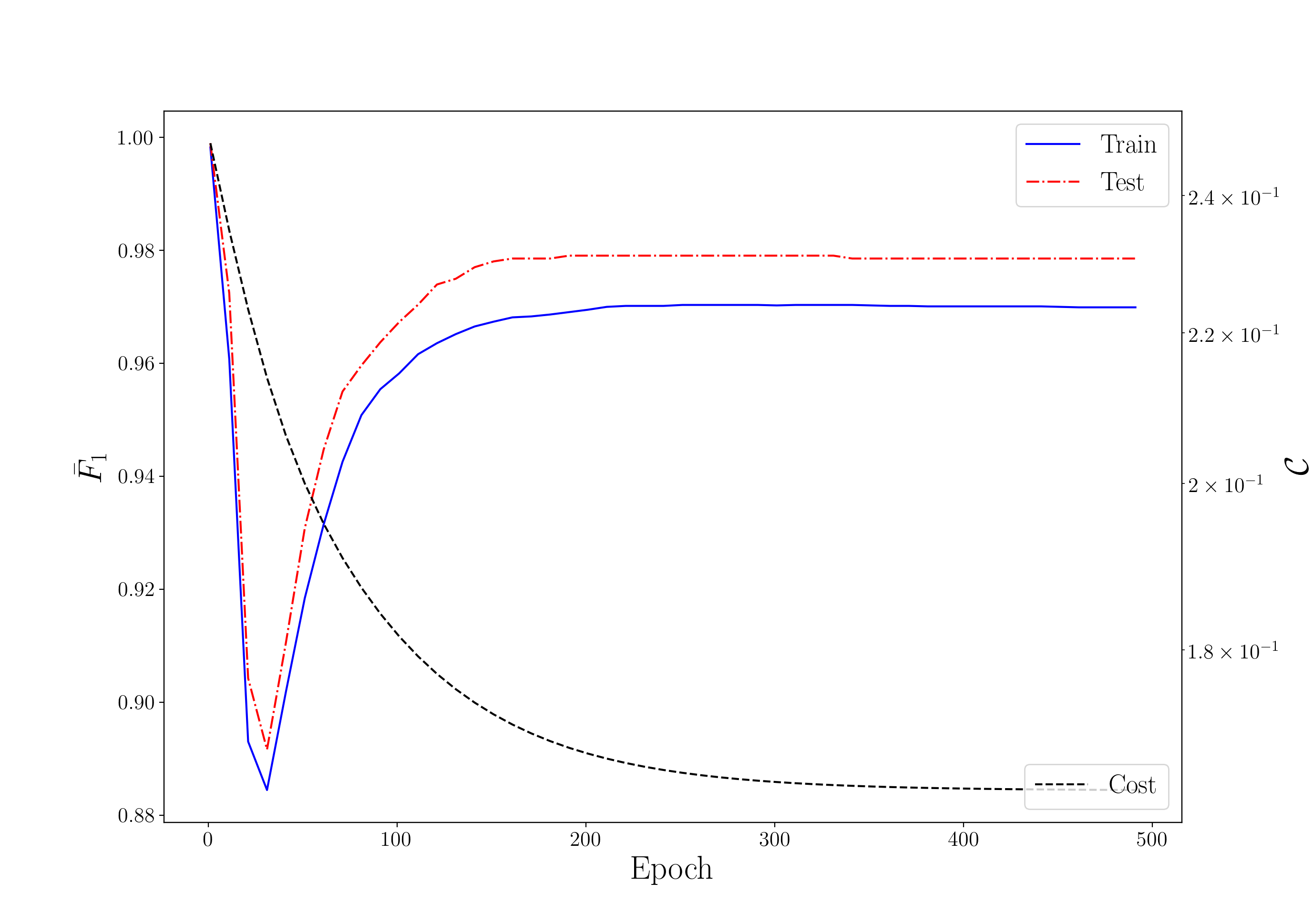}  
\caption{\label{fig:F1andC} (Color online) \emph{Evolution of $\bar{F}_1$ score and cost function during isometric optimization.} The training (blue solid) and test (red dot-dashed) average $F_1$-scores (left axis) and the cost function $\mathcal{C}$ from Eq.~\eqref{eq:quadcost} (black dashed, right axis) during the isometric optimization process of Eqs.~\eqref{eq:MGD1}-\eqref{eq:MGD3} with $\beta=0.1$ and $\eta=0.1$.  The starting point for optimization is the nearest isometric matrix to the unconstrained optimum, defined in Eq.~\eqref{eq:Wisonearest}.}
\end{center}
\end{figure}

We now investigate the interpretability of the decision weight vectors using the methodology developed in Sec.~\ref{sec:interp}.  For the purposes of having a square number of feature vectors for visualization, we use a cutoff of $\varepsilon=2\times10^{-4}$ on all layers up to the last, in which a bond dimension of 81 is used.  In Fig.~\ref{fig:NII}(a) we show the 81 feature vectors that form the basis for the decision space at the highest level of scale, visualized by fine-graining to the data scale, taking one-point averages, and displaying as an image.  Here, we see that the unsupervised feature extraction procedure results in feature vectors whose character is a mixture of the zero and one digits, with some vectors having a more zero-like character and others having a more one-like character.  Using the particular linear combination of feature vectors specified by the unconstrained weight vectors Eq.~\eqref{eq:W} for the 0 class and 1 class, we find the one-point averages shown in Figs.~\ref{fig:NII}(b) and (c), respectively.  Visually, we see little difference between the two vectors, with both having features of zero-like and one-like character.  To see how the two weight vectors define a classification boundary, we examine the difference of their one-point averages on a zoomed scale in panel (d).  We observe that the weight vector for class 0 is slightly larger in a ring surrounding the central pixels, while the weight vector for class 1 is largest on the central pixels.  The center pixel, in particular, has the largest difference for the two classes, indicating that this particular pixel is key for separating the two digits with the given weight vectors.

In the above we have utilized the ideal weight vectors that minimize the quadratic cost function Eq.~\eqref{eq:quadcost} defining our supervised classification problem.  We now look at optimizing an isometric set of weight vectors that define a mapping from the renormalized feature vectors into a class decision qubit.  In Fig.~\ref{fig:F1andC} we show the convergence behavior of the cost function during the manifold gradient descent procedure [Eqs.~\eqref{eq:MGD1}-\eqref{eq:MGD3}], starting from the nearest isometric matrix defined in Eq.~\eqref{eq:Wisonearest} and utilizing the parameters given in the figure.  As a point of reference, the unconstrained, non-isometric weights had a cost function of $\sim 4\times 10^{-6}$.  Also while the unconstrained optimization resulted in training and test $F_1$ scores of $0.99861$ and $0.99853$, the nearest isometric weights have scores of 0.99823 and 0.99897 and the results following 500 epochs of isometric optimization with the parameters in Fig.~\ref{fig:F1andC} are 0.96992 and 0.97856.  As expected, the cost function shows a monotonic decrease during optimization, but the $F_1$ scores show a non-monotonic behavior.

\begin{figure}[t]
  \begin{center}
\includegraphics[width=0.85\columnwidth]{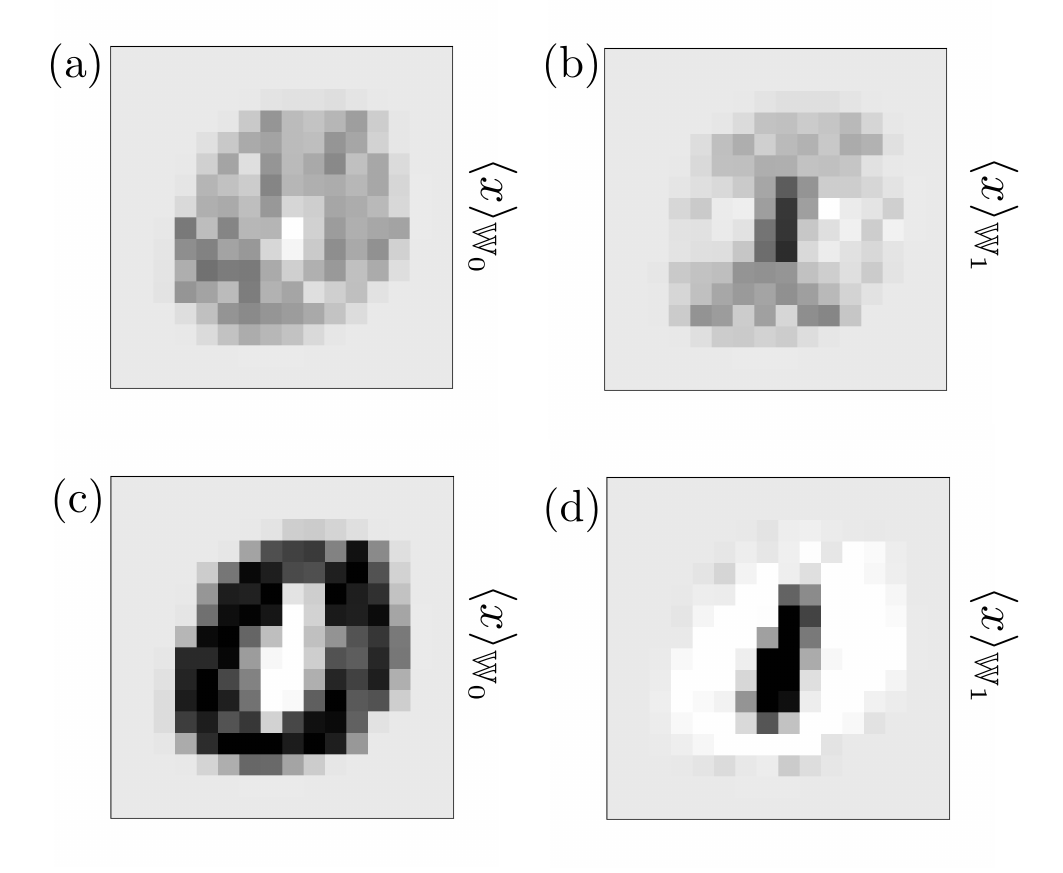}  
\caption{\label{fig:IsoInterp}  (Color online) \emph{One-point averages of isometric classification weight vectors at the data scale}.  Panels (a) and (b) are the analogs of Fig.~\ref{fig:NII}(b) and (c), utilizing the nearest isometric weight vectors to the unconstrained optimum, see Eq.~\eqref{eq:Wisonearest}.  Panels (c) and (d) are corresponding analogs following 500 epochs of manifold gradient descent as in Fig.~\ref{fig:F1andC}.  As gradient descent progresses, the corresponding data-scale averages become further distinguishable.}
\end{center}
\end{figure}

With the isometric weights in hand, we now re-visit the interpretability analysis from above, starting with a weight vector and fine-graining back to the data scale.  We note that the choice of weight vectors does not affect the interpretability of the coarse-grained feature vectors shown in Fig.~\ref{fig:NII}(a), but only their weighting for the final class decision.  The data-scale representation of the one-point averages of the isometric weight vectors are shown in Fig.~\ref{fig:IsoInterp}, with the upper row corresponding to the nearest isometric weight vectors to the unconstrained optimum, Eq.~\eqref{eq:Wisonearest}, and the lower row corresponding to the weight vectors following 500 epochs of manifold gradient descent with the parameters of Fig.~\ref{fig:F1andC}.  Comparing with Fig.~\ref{fig:NII} we can clearly see that the orthogonality constraint has resulted in more human-interpretable features, and has spread more of the decision importance throughout the data array rather than concentrating it into a few pixels, with both characteristics becoming more prominent following manifold optimization.  This motivates the use of isometric weights as a form of model regularization that may make the results less susceptible to adversarial perturbations, a key consideration for both classical~\cite{goodfellow2014explaining,chakraborty2018adversarial} and quantum~\cite{lu2020quantum,liu2020vulnerability} machine learning.  The use of orthogonality constraints in classical deep learning architectures has also been explored, see, e.g., Ref.~\cite{bansal2018can}.

\begin{figure}[t]
  \begin{center}
\includegraphics[width=0.85\columnwidth]{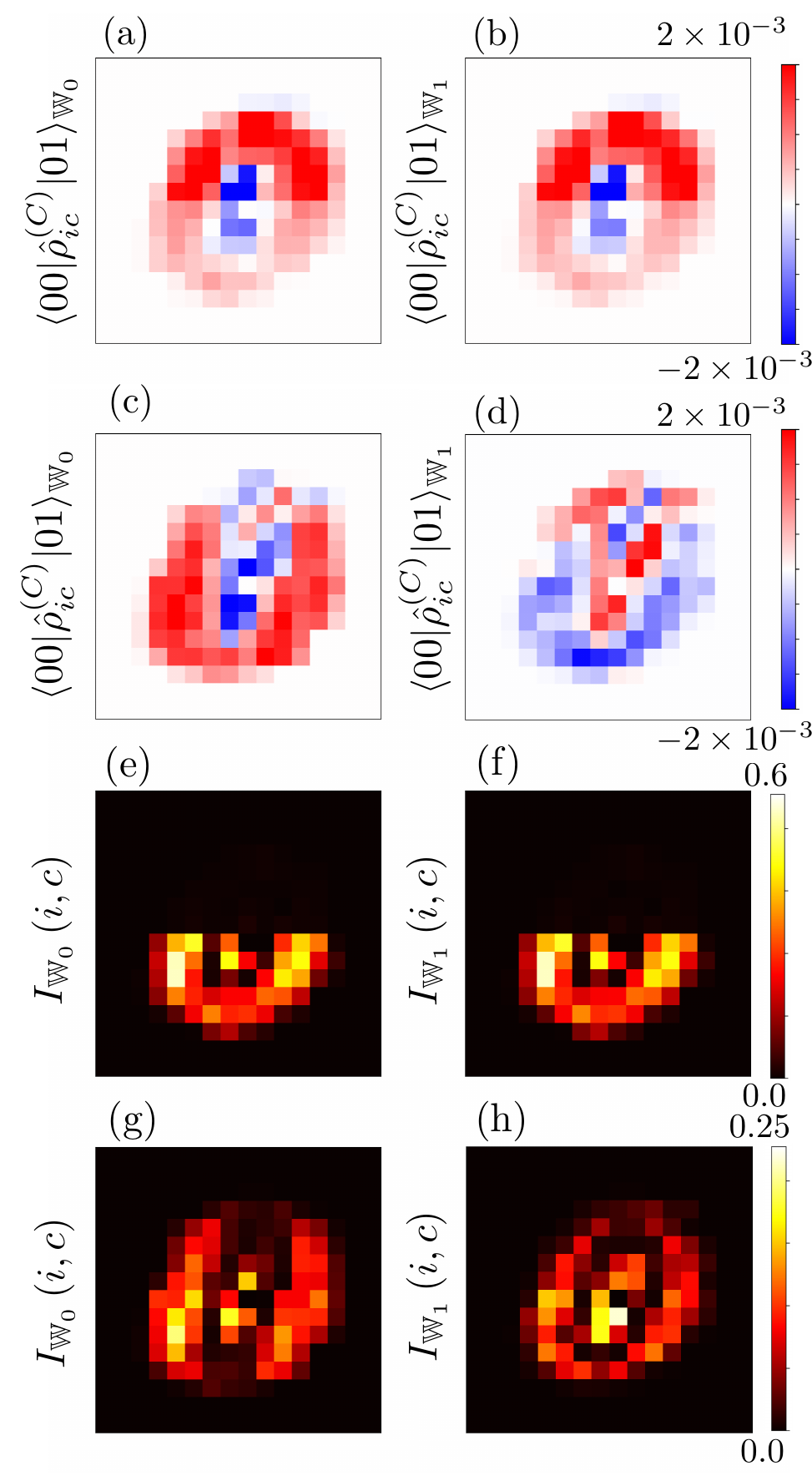}  
\caption{\label{fig:IsoInterp2}  (Color online) \emph{Two-point correlations of classification weight vectors at the data scale}.  Panels (a) and (b) are the elements $\langle 00|\hat{\rho}^{(C)}_{ci}|01\rangle$ of the correlation density matrix obtained from the unconstrained classification weights for the classes 0 and 1, respectively.  Panels (c) and (d) are the analogs for the isometric classification weights following gradient descent.  Panels (e) and (f) are the mutual information of the unconstrained weight vectors for class 0 and 1, respectively, evaluated between the center pixel and each other pixel in the array.  Panels (g) and (h) are the analogs for the isometric weight vectors following gradient descent.  The use of isometric data structures separates the two class decision vectors in the data space compared with the non-isometric case, and better utilizes correlations at the highest level of scale, resulting in more uniform values in the top and bottom halves of the arrays.}
\end{center}
\end{figure}

As mentioned in Sec.~\ref{sec:interp}, our interpretability metrics are not limited to one-point averages.  In Fig.~\ref{fig:IsoInterp2} we investigate two metrics of two-point correlations extracted from the weight vectors fine-grained to the data scale.  Here, we fix the location of one of the pixels to be the central location $c$, and display correlation metrics as a function of the other pixel location as an image.  Panels (a)-(d) use $\langle 00|\hat{\rho}^{(C)}_{ci}|01\rangle$, a coherence of the correlation density matrix between pixels $c$ and $i$ (Eq.~\eqref{eq:CDM}), which is a measure of correlation when positive and anticorrelation when negative.  Panels (a) and (b) are for the unconstrained weight vectors following normalization.  Similar to Fig.~\ref{fig:NII}, very little distinction is seen between the data-scale metric for the class 0 and class 1 weight vectors.    Another clear characteristic of these panels is that the correlation is different for the upper and lower half of the image.  Given that the final tensor in the feature extractor encodes correlations between the upper and lower halves of the pixel array (see Fig.~\ref{fig:ResampTopo}), this indicates that the weight vectors are not significantly utilizing correlations at the highest level of scale.  We can contrast this with panels (c) and (d), which are the same metrics for the isometric weight vectors following gradient descent.  These correlations show the same order of magnitude on the top and bottom of the pixel array, demonstrating that correlations are weighted similarly at all scales, and we see clear patterns of correlation and anticorrelation that are different between the class 0 and class 1 weight vectors.  Panels (e)-(h) use the metric of mutual information (Eq.~\eqref{eq:MI}), with the upper two panels being the (normalized) unconstrained weight vectors and the lower two panels being the isometric weight vectors following gradient descent.  Similarly to the correlation density matrix metric, the unconstrained weight vectors show no visible difference between class 0 and class 1, and show a higher degree of correlation in the lower half of the pixel array (containing the fixed pixel location $c$) than in the upper half.  The isometric weight vectors in the lower two panels show similar order of magnitude of mutual information in the upper and lower halves of the pixel array, and distinct patterns for the two classes.

\subsection{Human activity recognition}

In order to demonstrate applicability of TTN-based classifiers to multivariate time series data, we also apply the above methods to a dataset for human activity recognition (HAR)~\cite{anguita2013public}.  The dataset is formed from smartphone accelerometer and gyroscope data recorded for 15 second intervals at 50Hz from 30 subjects.  The activities comprise six classes: walking, walking upstairs, walking downstairs, sitting, standing, and laying.  The data were further processed using a median filter and a 3rd-order low-pass filter with 20Hz cutoff frequency to reduce noise, and an additional low-pass filter with a cutoff frequency of 0.3Hz was utilized to remove gravitational forces from the accelerometer data.  Finally, the data was re-sampled in sliding windows of 2.56 seconds with 50\% overlap to result in 128 data points per timeseries.  The training (test) dataset consists of 7352 (2947) collections of timeseries.

\begin{figure}[t]
  \begin{center}
\includegraphics[width=0.99\columnwidth]{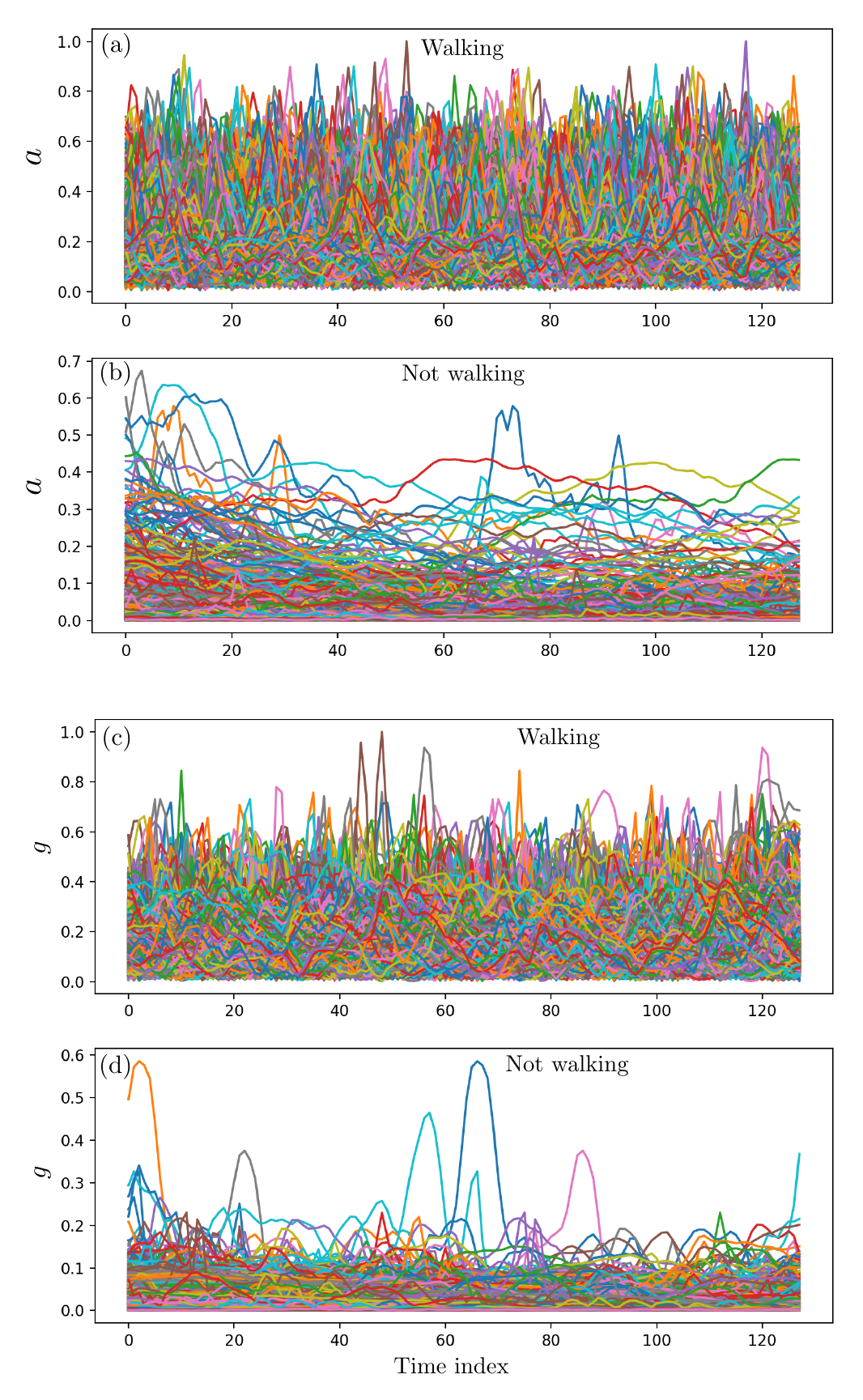}  
\caption{\label{fig:agtrain}  (Color online) \emph{Derived time series features for the HAR dataset}. The dimensionless rescaled acceleration vector norms [class 0, panel (a) and class 1, panel (b)] and angular velocity vector norms [class 0, panel (c) and class 1, panel (d)] for the data in the training dataset are displayed as a function of the time index.}
\end{center}
\end{figure}

We will focus on the task of classifying a time series as being either walking (i.e., coming from the walking, walking upstairs, or walking downstairs classes of the original dataset), which we will refer to as class 0, or not walking (coming from the sitting, standing, or laying classes of the original dataset), which we will refer to as class 1.  We then utilize the norms of the total acceleration vector $a\left(t\right)=\sqrt{a_x^2\left(t\right)+a_y^2\left(t\right)+a_z^2\left(t\right)}$ and total angular velocity vector $g\left(t\right)=\sqrt{\omega_x^2\left(t\right)+\omega_y^2\left(t\right)+\omega_z^2\left(t\right)}$ as input features to the TTN feature extractor.  The final stage in processing is to rescale all data to the interval $a\left(t\right)\in\left[0,1\right]$, $g\left(t\right)\in\left[0,1\right]$ $\forall t$, which also renders the data dimensionless.  The data in the training set, processed as described above, are shown for the two classes in Fig.~\ref{fig:agtrain}, with the upper panels displaying the timeseries for $a\left(t\right)$ and the lower panels displaying the timeseries for $g\left(t\right)$.

\begin{figure*}[t]
  \begin{center}
\includegraphics[width=1.9\columnwidth]{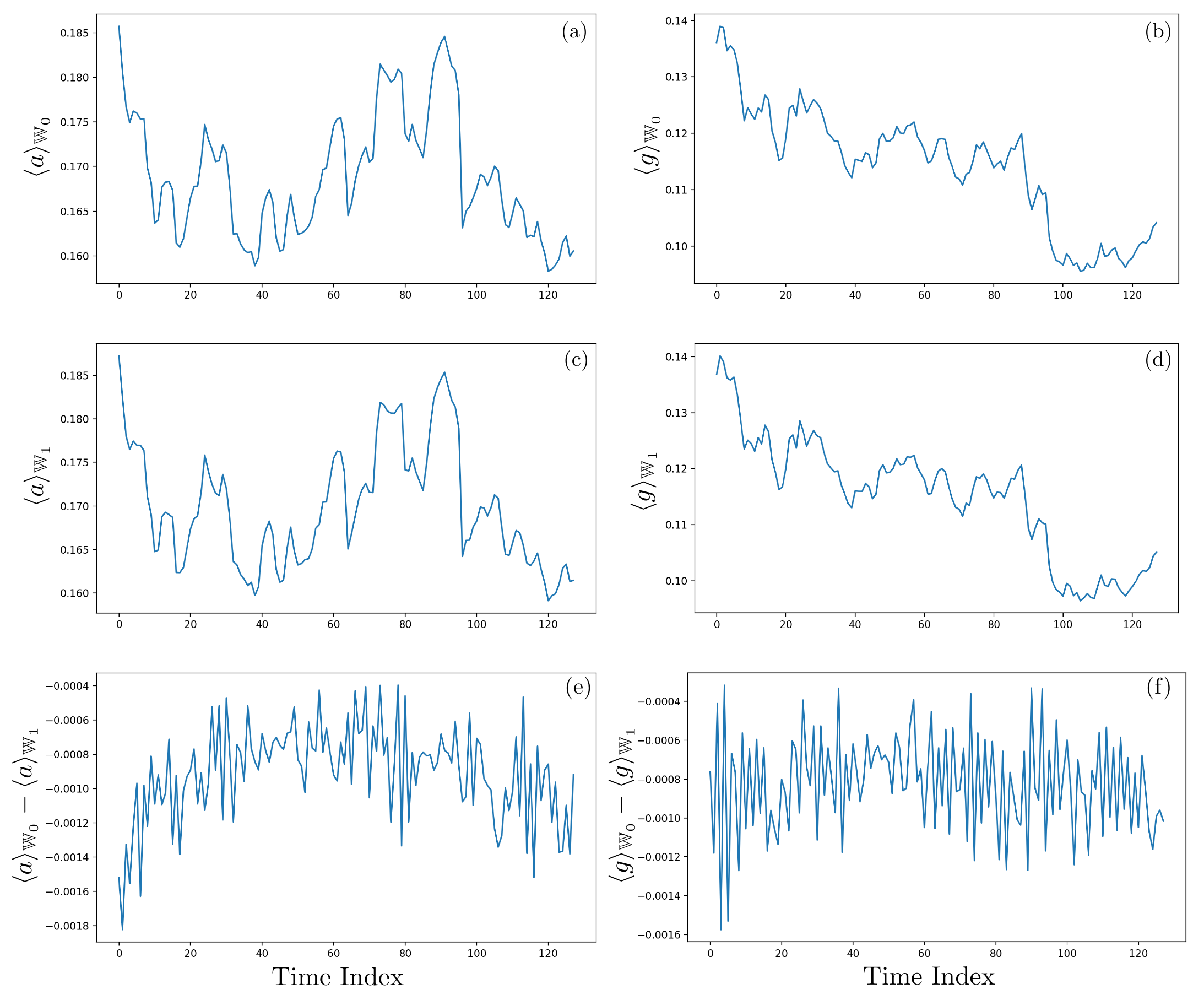} 
\caption{\label{fig:agInterp}  (Color online) \emph{Average one-point representation of classification weight vectors at the data scale}.  The data-scale representations of the class decision weight vectors are shown for class zero (one) in the top (middle row), with their difference in the bottom row.  The left column is the acceleration vector norm feature, and the right column is the angular velocity vector magnitude feature.  Analogous to the MNIST example in Fig.~\ref{fig:NII}, very little difference is discernible between the two classes from the one-point averages.}
\end{center}
\end{figure*}

We apply a TTN feature extractor to the full time series, with the data ordering $\left[a\left(1\right),g\left(1\right),a\left(2\right),g\left(2\right),\dots,a\left(L\right),g\left(L\right)\right]$ such that the first layer mediates correlations $a$ and $g$ at the same time, and higher layers facilitate correlations between all features at different times.  Using this topology, a cutoff of $\varepsilon=1\times10^{-6}$, and a maximum bond dimension of 812, we find training and test average $\bar{F}_1$ scores of 0.9974 and 0.9627, respectively, with the weight vectors found by unconstrained optimization, Eq.~\eqref{eq:W}.  In Fig.~\ref{fig:agInterp} we show the results of our average one-point interpretability analysis, Eq.~\eqref{eq:xinterp}, for the class weight vectors in the two classes and for the two features $a$ and $g$, together with the differences between the time series for the two cases.  Similar to what was seen in the example using MNIST image data shown in Fig.~\ref{fig:NII}, very little difference can be seen between the one-point averages at the data scale, here interpreted as time series.  In contrast to the MNIST example, in which a more detailed analysis of the difference of the one-point averages identified data regions that were critical to the class decision, in the present case the difference of the time series displays significant variation across the range of the data and there is not an immediate interpretation of the data used in forming the class decision.

\begin{figure*}[t]
  \begin{center}
\includegraphics[width=1.99\columnwidth]{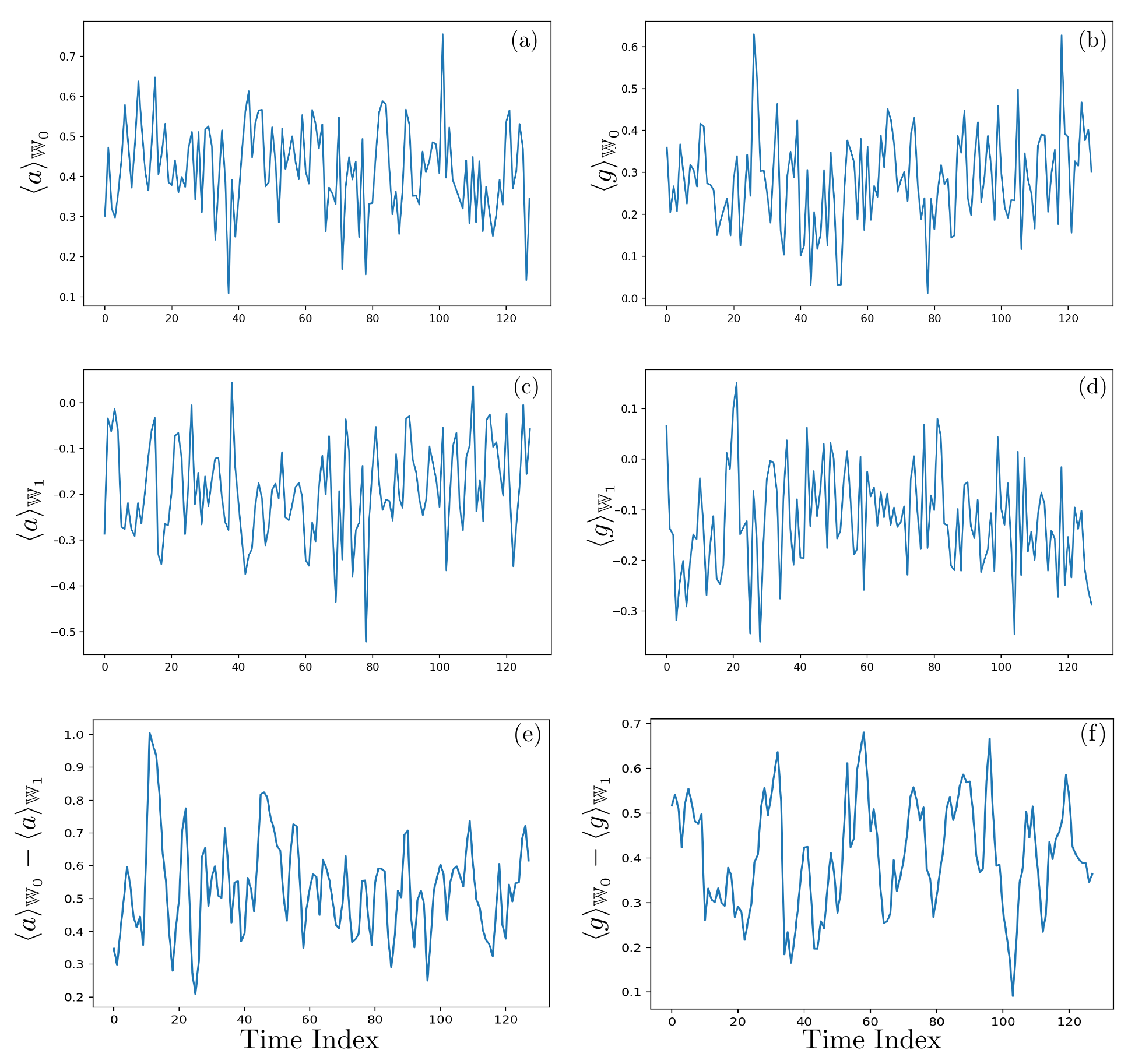} 
\caption{\label{fig:agInterpiso}  (Color online) \emph{Average one-point representation of isometric classification weight vectors at the data scale}.  The data-scale representations of the isometric class decision weight vectors are shown for class zero (one) in the top (middle row), with their difference in the bottom row.  The left column is the acceleration vector norm feature, and the right column is the angular velocity vector magnitude feature.  The isometric constraint separates the time series for the two classes significantly compared with the unconstrained case in Fig.~\ref{fig:agInterp}.}
\end{center}
\end{figure*}

We now investigate the case in which the class decision weight vectors are restricted to form an isometric matrix, obtained by manifold gradient descent.  We find a final cost function of $0.1729$, which should be compared with the cost function $6.06\times10^{-3}$ found by unconstrained optimization, following $\sim 1000$ epochs of manifold gradient descent with a final gradient norm of $\sim 4\times 10^{-5}$.  The final training and test average $\bar{F}_1$ scores are 0.9418 and 0.9983, respectively.  The difference in performance is almost completely due to data elements from class zero being misclassified as class one; the training dataset has 400 misclassified elements of class zero while the test dataset properly classifies all elements from class zero.  With the isometric class weight vectors in hand, we now repeat the one-point interpretability analysis shown in Fig.~\ref{fig:agInterp}.  The results are shown in Fig.~\ref{fig:agInterpiso}.  An immediately striking feature of the interpretability analysis is that the one-point average time series for class 1 are negative, while the definition of the input features as norm vectors requires ``physical" features to be positive.  An interpretation of this can be traced back to Eq.~\eqref{eq:xinterp}, where the one-point average is extracted.  A negative value of the time series implies that the two amplitudes of the qubit encoding at the lowest level of scale have opposite signs.  To understand how this affects the class decision, we can define a ``mean-field" weight vector for class $c$ at the lowest level of scale as
\begin{align}
\label{eq:MF1}|\Phi_{\mathrm{MF};c}\rangle&=\otimes_{j=1}^{L}\left(\sum_{i_j=1}^{2} \phi_{i_j}^{(j)}\left(\langle x_j\rangle_{\mathbb{W}_c}\right)|i_j\rangle\right)\, .
\end{align}
The dot product with a data vector $\mathbf{z}$ at this level of scale is
\begin{align}
\langle \Phi^{\left(0\right)}\left(\mathbf{z}\right)|\Phi_{\mathrm{MF};c}\rangle&=\prod_{j=1}^{L}\boldsymbol{\phi}^{(j)}\left(z_j\right)\cdot \boldsymbol{\phi}^{(j)}\left(\langle x_j\rangle_{\mathbb{W}_c}\right)\\
&=\prod_{j=1}^{L}\cos\left[\frac{a}{x_{\mathrm{max}}}\left(z_j-\langle x_j\rangle_{\mathbb{W}_c}\right)\right]\, .
\end{align}
Recalling that $a\frac{x}{x_{\mathrm{max}}}\ll 1$, each of the cosines can be expanded as
\begin{align}
\label{eq:MFN}\cos\left[\frac{a}{x_{\mathrm{max}}}\left(z_j-\langle x_j\rangle_{\mathbb{W}_c}\right)\right]\approx 1-\frac{a^2}{2x_{\mathrm{max}}^2}\left(z_j-\langle x_j\rangle_{\mathbb{W}_c}\right)^2\, .
\end{align}
Hence, the contribution from the component of test data vector $\mathbf{z}$ at time point $j$, $z_j$, to the class decision is given roughly by its distance from the average time series vectors for the two classes at this time point $\langle x_j\rangle_{\mathbb{W}_c}$.  While the test data elements $z_j$ will never take on negative values, they may be closer to small negative values than large positive values, and the appearance of negative values in $\langle x_j\rangle_{\mathbb{W}_c}$ may facilitate orthonormality between the weight vectors in the various classes.

\begin{figure*}[t]
  \begin{center}
\includegraphics[width=1.99\columnwidth]{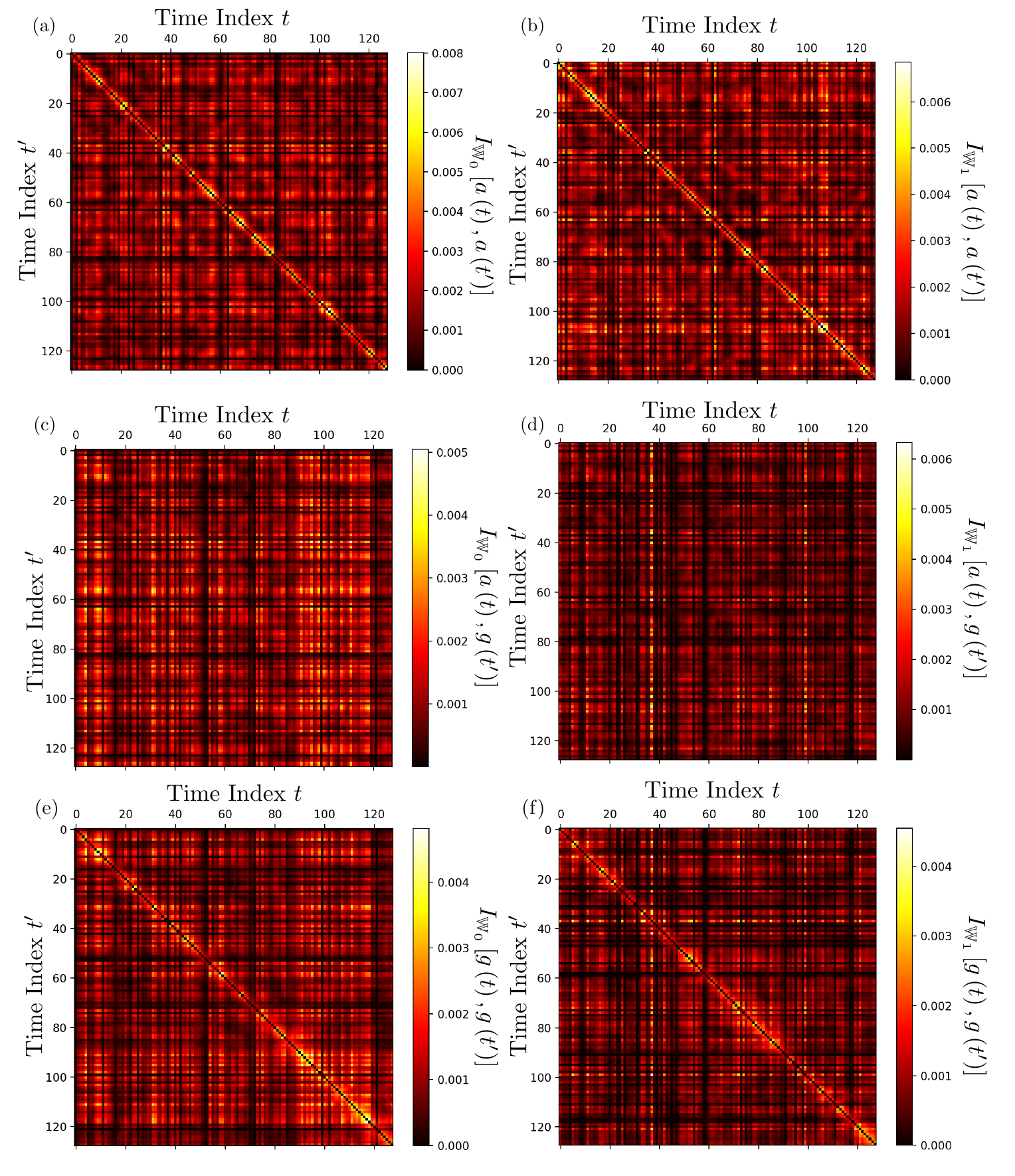} 
\caption{\label{fig:agMIiso}  (Color online) \emph{Two-point mutual information of isometric classification weight vectors at the data scale}.   The mutual information metric Eq.~\eqref{eq:MI} of the isometric classification decision weight vectors for class zero (left column) and class one (right column) are shown as functions of time index.  The top row displays correlations between acceleration features at different times, the middle row gives cross-correlations between accelerations and angular velocities at different times, and the bottom row shows correlations between angular velocities at different times.  In all cases, correlations are seen between features at all timescales, with the strongest correlations seen between acceleration features at nearby times.}
\end{center}
\end{figure*}

The above one-point analysis in Eq.~\eqref{eq:MF1}-\eqref{eq:MFN} does not account for correlations between the $a$ and $g$ features, or between features at different times, and so does not fully capture all of the information present in a correlated quantum model.  To look at correlations beyond one-point averages, we look at the two-point mutual information between acceleration and angular velocity features at different times, displayed in Fig.~\ref{fig:agMIiso}.  Here, the left (right) column is the mutual information of the classification decision weight vector for class zero (class one), and the top row, middle row, and bottom row are correlations between accelerations, cross-correlations between accelerations and angular velocities, and correlations between angular velocities at different times, respectively.  The weight vectors display a complex set of correlation behaviors, with correlations seen between all features spread across timescales at comparable magnitudes.  However, we do find that the strongest correlations are those between the acceleration features at nearby times.

\section{Conclusions and outlook}

We have investigated classifiers based on an encoding of classical data vectors into quantum states in a Hilbert space that is exponentially large in the length of the data vector.  An unsupervised feature extractor with a tree tensor network (TTN) topology extracts a relatively small basis of relevant quantum states from a training set of data embedded into quantum states, with the size of this basis being the main hyperparameter of the model.  The tensors in this network are optimized utilizing a procedure which keeps the dominant correlations between the quantum degrees of freedom encoding elements of the data vector, analogous to a renormalization group flow, at increasingly coarse levels of scale.  The extracted feature vectors at the highest level of scale were utilized in a supervised cost function optimization to define a classification decision.  We presented novel metrics for interpretability of such quantum classifiers which extract low-order correlations from the weight vectors interpreted as a quantum state and fine-grained to the data scale by running the feature extraction network in reverse.  In contrast to previous work on quantum-inspired algorithms using hierarchical tensor networks for classification, we utilized tools for the optimization of such models in which all elements correspond to proper quantum data structures that can be implemented on gate-based quantum computing devices.  This included devising an embedding map with learning properties comparable to well-performing classical embedding maps, but which also produces valid quantum states.  In addition, we utilized manifold-based optimization schemes that define an isometric mapping from the quantum feature vectors produced by the unsupervised feature extractor into a register of qubits whose probability amplitudes define a class decision.  We discussed methods for translating the isometric tensors obtained through classical optimization into operations to be performed on a gate-based quantum computer, and also discussed scaling of quantum resource requirements.

We demonstrated the methods developed in this work on two datasets: the canonical MNIST handwritten digit dataset and a multivariate time series dataset of human activity recognition (HAR).  We demonstrated that a small-angle phase embedding map, which produces valid quantum states from classical data, gives comparable performance on the full MNIST dataset to the encoding of Ref.~\cite{novikov2016exponential} mapping the data to a high-order polynomial.  We then compared the results of a simpler classification problem of distinguishing the digits zero and one from the MNIST dataset using a unconstrained, quantum-inspired model with those from the quantum model encoding the classification decision into the amplitudes of a qubit.  We find that the use of fully quantum data structures produces more human interpretable features, and utilizes more of the information and correlations across the data vector in making a classification decision, potentially improving robustness against noise or adversarial perturbations.  Similar qualitative behavior was seen in the case of time series from the HAR dataset, where we demonstrated the applicability of TTN-based classifiers to large-dimensional multivariate time series.

Several opportunities exist for further research in quantum-assisted machine learning using tensor networks with a TTN structure.  For one, the tensors in our TTN feature extractor were defined using an unsupervised approach, but better performance may be had by using the isometric optimization methods described in this paper to the tensors of the feature extractor.  More complex structures beyond trees, such as the multiscale entanglement renormalization ansatz (MERA) network~\cite{PhysRevLett.101.110501} or combinations of trees and linear networks such as matrix product states~\cite{stoudenmire2018learning,reyes2020multi}, can be considered, and can also be optimized using manifold gradient descent.  Investigations of the fidelity of TTN models compiled to currently available quantum hardware and their resilience to noise, analogous to Ref.~\cite{wall2020Generative} for the case of generative matrix product state models, can be performed.  Alternative training strategies, e.g. those that utilize partial data vectors or regularization terms, may provide better performance than the simple cost function utilized here.  Finally, it is intriguing to consider methods for the optimization of TTN-based classifiers with a quantum device in the loop, where performance can be investigated as the network is scale to the classically intractable regime. 

\section{Acknowledgements}

We would like to thank Matt Abernathy and Greg Quiroz for useful discussions, and would like to acknowledge funding from the Internal Research and Development program of the Johns Hopkins University Applied Physics Laboratory.

\bibliography{Refs}
\bibliographystyle{apsrev4-1}

\end{document}